\documentclass[aps,pre,amsmath,twocolumn,superscriptaddress,showpacs]{revtex4-1}
\usepackage{fancyhdr}
\usepackage{calc}
\usepackage{amsmath}
\usepackage{amsfonts}
\usepackage{amssymb}
\usepackage{graphicx}
\usepackage{soul}
\usepackage{mathtools}
\usepackage{verbatim}
\usepackage{epstopdf}
\usepackage{subfigure}
\usepackage{latexsym}
\usepackage{dcolumn}
\usepackage{epsf}
\usepackage{float}
\usepackage[table]{xcolor}
\usepackage{multirow}
\usepackage{color} 
\usepackage{graphicx}% Include figure files
\usepackage{graphicx,epstopdf}
\usepackage{xcolor}
\usepackage{dcolumn}% Align table columns on decimal point
\usepackage{bm}% bold math
\usepackage{mathrsfs} % script-like, curvy letters.
\usepackage{mathrsfs}
\usepackage{url}
\usepackage{hyperref}
\usepackage{breakurl}

\DeclareMathOperator\erf{erf}

\begin{document}
\title{Consensus Formation Among Mobile Agents in Networks of Heterogeneous Interaction Venues}%Locale}% Venues}%Environments}%Impact of network topology on consensus among mobile agents}

\author{Guram Mikaberidze}
%\email{mikaberidze@ucdavis.edu}
\affiliation{Department of Mathematics, University of California, Davis, CA, 95616, USA}
\email{mikaberidze@ucdavis.edu}

\author{Sayantan Nag Chowdhury}
%\email{jcjeetchowdhury1@gmail.com}
\affiliation{Department of Environmental Science and Policy, University of California, Davis, CA 95616, USA}
\email{jcjeetchowdhury1@gmail.com}

%\author{***************}
\author{Alan Hastings}
\affiliation{Department of Environmental Science and Policy, University of California, Davis, CA 95616, USA}
\affiliation{Santa Fe Institute, 1399 Hyde Park Road, Santa Fe, NM 87501, USA}	
\email{amhastings@ucdavis.edu}

%\author{***********}
\author{Raissa M. D’Souza}
\affiliation{Department of Computer Science and Department of Mechanical and Aerospace Engineering, University of California, Davis, CA 95616, USA}
\affiliation{Santa Fe Institute, 1399 Hyde Park Road, Santa Fe, NM 87501, USA}
\affiliation{Complexity Sciences Hub Vienna, Vienna, Austria}
\email{raissa.cse@gmail.com}

%\subject{epidemic spreading, complex network}

%\keywords{Complex network, epidemic spreading, final outbreak size, test-kit}
%\thanks{These two authors contributed equally}
	
%\thanks{Corresponding Auhtor: Chittaranjan Hens}
%\email{chittaranjanhens@gmail.com}

%%%% Abstract text to be placed here %%%%%%%%%%%%
\date{\today}
\begin{abstract}
Exploring the collective behavior of interacting entities is of great interest and importance. Rather than focusing on static and uniform connections, we examine the co-evolution of diverse mobile agents experiencing varying interactions across both space and time. Analogous to the social dynamics of intrinsically diverse individuals who navigate between and interact within various physical or digital locations, agents in our model traverse a complex network of heterogeneous environments and engage with everyone they encounter. The precise nature of agents' internal dynamics and the various interactions that nodes induce are left unspecified and can be tailored to suit the requirements of individual applications. We derive effective dynamical equations for agent states  which are instrumental in investigating thresholds of consensus, devising effective attack strategies to hinder coherence, and designing optimal network structures with inherent node variations in mind. We demonstrate that agent cohesion can be promoted by increasing agent density, introducing network heterogeneity, and intelligently designing the network structure, aligning node degrees with the corresponding interaction strengths they facilitate. Our findings are applied to two distinct scenarios: the synchronization of brain activities between interacting individuals, as observed in recent collective MRI scans, and the emergence of consensus in a cusp catastrophe model of opinion dynamics.
\end{abstract}


\maketitle 

\section{Introduction}

In recent years, the scientific community has made impressive progress in comprehending the complexities of interacting systems. The methodology of network science \cite{barabasi2016network} has provided a powerful framework for this quest. It has unveiled fresh insights into the role of the network structure, wielding profound influence over collective behavior \cite{dorogovtsev2008critical,barrat2008dynamical}, in diverse realms spanning from the networks of cortical neurons to the fabric of society. However, most studies have focused on scenarios where the interactions between units remain constant over time, neglecting numerous realistic situations with time-varying interactions \cite{holme2012temporal}, such as person-to-person communication \cite{iribarren2009impact,wu2010evidence}, cooperative dynamics of animal groups \cite{liu2013contagion} and rational individuals \cite{nag2020cooperation}, and robot and vehicle movements \cite{bullo2009distributed,sun2016velocity}, among others.

% Over the past decade, the scientific community has made incredible progress in comprehending the complexities of interacting systems. The methodology of network science \cite{barabasi2016network}, which delves into the interactions between the constituents of these complex systems, has provided a powerful framework for this quest. It has unveiled fresh insights into the interplay between the network's structure and its dynamical behavior in diverse realms, spanning from the networks of cortical neurons to the fabric of society itself. Nowadays, it is widely recognized that the underlying network's topology wields profound influence over collective behavior \cite{dorogovtsev2008critical,barrat2008dynamical}.
%
%Nevertheless, most studies have focused on scenarios where the interactions between units remain constant over time. In reality, numerous realistic situations exist where interactions among units vary with time \cite{holme2012temporal}, such as person-to-person communication \cite{iribarren2009impact,wu2010evidence}, cooperative dynamics of animal groups \cite{liu2013contagion} and rational individuals \cite{nag2020cooperation}, and robot and vehicle movements \cite{bullo2009distributed,sun2016velocity}, among others. The mounting evidence underscores the urgent need to broaden our perspective and move away from the assumption of static and unchanging interactions within networks. 

\begin{figure}[h]
	\centerline{\includegraphics[scale=0.5]{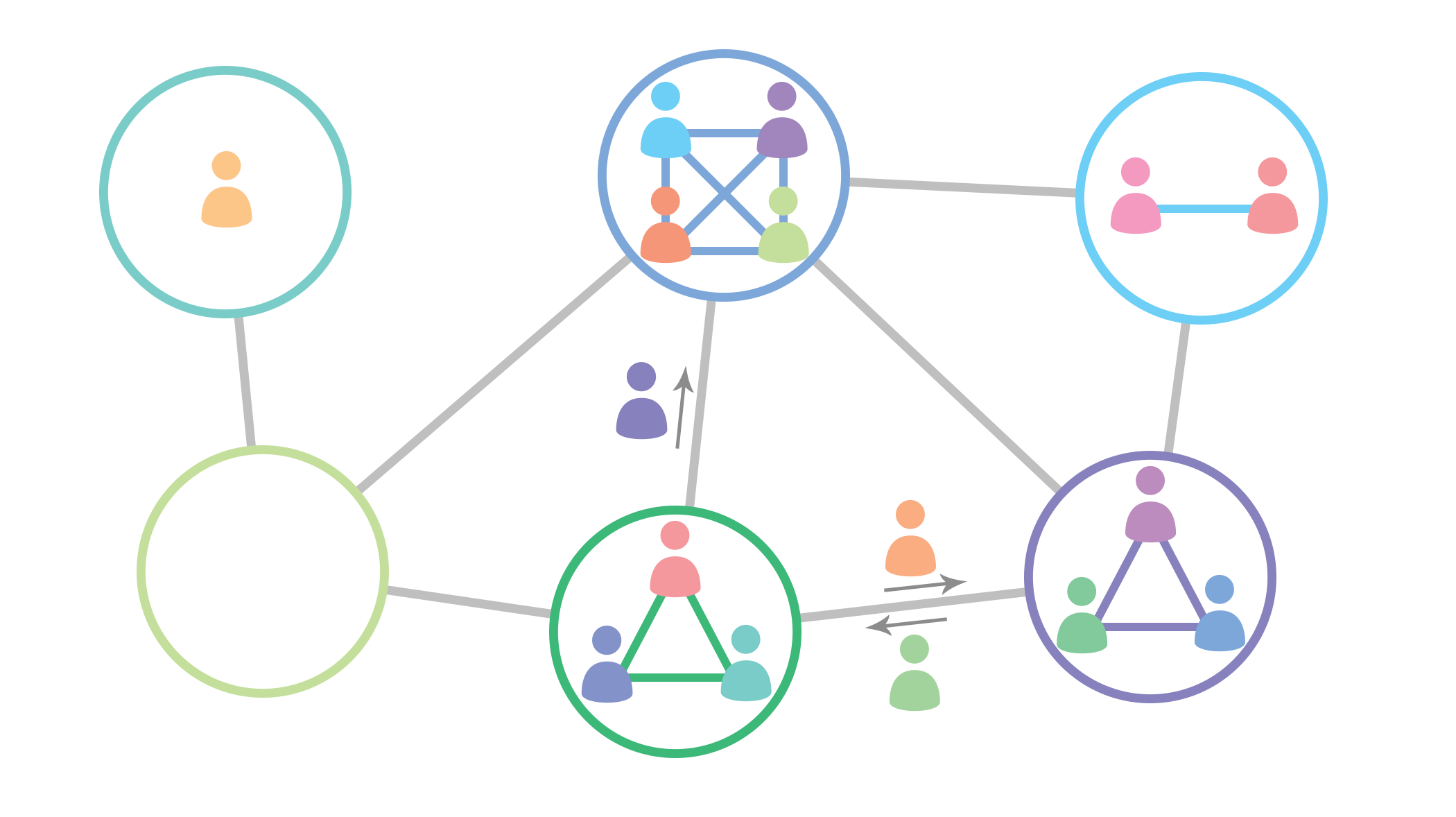}}
	\caption{
		Network of heterogeneous locations hosting diverse mobile agents. Agents move around the network and interact with other agents. Types of agent interactions depend on the host.
	}\label{illustration}
\end{figure}

Here, we study the \textit{co-evolution of diverse mobile agents with interactions varying across space and time}. Consider how social consensus emerges when individuals, each with a unique thinking pattern, navigate between and interact within varied physical or digital locations. Accordingly, in our model, various agents navigate a complex network of diverse locations, interacting with everyone they meet along the way (see Fig.\ \ref{illustration}). To maintain realism, we assume that nodes can facilitate a range of interactions, while the specific forms of agents’ internal dynamics and interactions are deliberately left unspecified and can be chosen based on the application. Our approach yields concise effective dynamical equations for agent states in the weak coupling limit. These equations serve as critical tools to explore thresholds of consensus, crucial in opinion dynamics, devise effective strategies to hinder coherence, and design optimal network structures with inherent node variations in mind.

We find that enhancing the effective interaction strength and thus promoting coherence can be achieved by: increasing the number of agents, reducing the network size, aligning node degrees with the interactions they induce, or augmenting the degree distribution in the network. The latter serves as a prime example of converse symmetry breaking \cite{nishikawa2016symmetric}, since discrepancies in node degrees promote unity among agent states. We also find that a strategic approach to disrupt coherence involves targeting high-degree nodes due to their extensive influence on collective behavior. %Indeed, the density-dependent synchronization threshold, as determined through our calculations, is also evident in various phenomena such as bacterial infection, biofilm formation, and bioluminescence, unveiling quorum-sensing transitions in coupled systems \cite{nadell2008evolution,taylor2009dynamical,camilli2006bacterial,chowdhury2019synchronization20}.

For validation and applications, we will introduce specific internal dynamics and interactions for agents. As the first example, we will delve into an intriguing line of experimental research that examines the ``conceptual alignment" or ``brain-to-brain synchronization" between interacting individuals \cite{wheatley2019beyond, hu2017brain, perez2017brain}. Such experiments often utilize collective MRI and EEG brain scans \cite{czeszumski2020hyperscanning}. This neuroimaging technique is called ``Hyperscanning" and it simultaneously records brain activity from multiple individuals during social interactions or coordinated tasks. Recent observations have shown that brain activity patterns in response to stimuli become synchronized and remain so after engaging in a common discussion \cite{sievers2020consensus}. Similarly, the pupils of conversing individuals contract and dilate in synchrony \cite{wohltjen2021eye}. To represent these interactions, we will model the participants as Kuramoto agents that synchronize during constructive discussions and desynchronize during disruptive interactions. This example is similar to the metapopulation model \cite{gomez2013motion} where Kuramoto agents perform a degree biased random walk on a network. However, in contrast from Ref. \cite{gomez2013motion}, our work considers non-uniform couplings and allows them to be repulsive.

As the second application, we will explore a mathematical model of polarization within and across individuals \cite{van2020polarization}. This model incorporates internal dynamics based on a cusp catastrophe of opinion, which is a function of external influence and individuals' attention to the subject matter. We will incorporate this model as the agents' internal dynamics and study the feasibility of consensus.

In both application systems, the individuals will navigate various social settings, including online platforms like the comments sections of news articles and social media posts, as well as in-person gatherings such as offices, schools, bars, and book clubs. Each interaction venue is considered a distinct node in the network. During these gatherings, agents engage in interactions with all other participants present in the same node. Experimental studies have shown that exposure to a different point of view can lead to either convergence \cite{balietti2021reducing} or divergence \cite{bail2018exposure} in opinions. It is plausible that the nature of the interaction setting plays an important role in this outcome. Thus, in our model, the network nodes represent diverse environments inducing various interactions among the visitors. For example, interactions in a bar are expected to be different from interactions in a debate club. Some nodes will foster consensus by introducing attractive (cohesive) interactions between interacting agents, while others may contribute to discord by imposing repulsive (disruptive) interactions.

The literature on mobile agents \cite{prignano2013tuning,frasca2012spatial,perez2017control,zhou2017connection,majhi2019emergence,buscarino2016interaction} remains limited and most studies have focused on two simplistic assumptions. First, they consider mobile agents moving randomly on a continuous two or three-dimensional plane. In contrast, we use a novel approach assuming that mobile agents traverse a complex network, hopping node to node along the edges. This movement exposes them to varying sets of neighbors. Second, the interactions in the literature are usually fixed and independent of the agents absolute location. Our model overcomes this assumption by allowing the locations to dictate how the agents will interact. The local interactions on the nodes promote local coherence of the internal dynamics whereas the random walk continuously updates the interacting subsets of agents, facilitating global coherence. It is important to note that our proposed model only partially reflects the complexities of real-world scenarios. Nonetheless, in an effort to capture the richness of natural settings, we incorporate both cohesive and disruptive interactions \cite{hong2011kuramoto,chowdhury2020effect,restrepo2006synchronization,chowdhury2021antiphase,jalan2019inhibition,chowdhury2023interlayer,hong2011conformists,chowdhury2020distance,leyva2006sparse,chowdhury2019synchronization}.
%When mobile agents visit a node with positive coupling, they tend to align their internal states, leading to local coherence. Conversely, when they visit a node with a negative coupling, they strive to maximize their differences, akin to an antiferromagnetic interaction. 
 The interplay between positive and negative couplings holds significant relevance in neuronal networks comprised of both excitatory and inhibitory neurons \cite{vogels2009gating,soriano2008development}. Similarly, in social networks, one can discern the coexistence of contrarians alongside conformists \cite{zhang2013efficient}, giving rise to starkly contrasting dynamics evident in phenomena such as political elections or the spread of rumors. The coexistence of these two types of couplings allows our model to capture a wide range of realistic settings \cite{majhi2020perspective}. By introducing these advancements to the study of mobile agents, we aim to expand our understanding of complex systems and their collective behaviors, acknowledging the inherent simplifications of our model while embracing its potential to capture essential aspects of real-world dynamics.

 With the aforementioned objectives in mind, we embark on addressing the following pivotal questions through analytical means:

\begin{enumerate}
	\item {\it Can coherence be achieved among mobile agents, even under disruptive influences? We aim to determine the critical threshold analytically, indicating the number and strength of disruptive  nodes required to fully disrupt coherence.}
	
	\item {\it How does the network topology impact the collective behavior, among these mobile agents when subjected to the combined influence of attractive and repulsive interactions? Can we discern which network topology is more robust in the face of repulsive interactions?}
	
\end{enumerate}

To unravel the answers to these inquiries, we first provide an in-depth exposition of our model in Sec.\ \ref{Mathematical Model}. In Sec.\ \ref{Mean-field analysis}, we give a comprehensive analytical derivation of our main result, the effective equations of agents' internal states. We also discuss the general insights offered by them. In Sec.\ \ref{Exploring Synchronization: Random walk of Kuramoto agents} we focus on the application example of Brain-to-Brain synchronization, and, for the first time in Sec.\ \ref{Kuramoto model as internal dynamics of mobile agents}, we introduce specific free evolution and interactions into our system. We also need to specify the distribution of disruptive and cohesive nodes. Section\ \ref{Untargeted attacks} considers the ``untargeted attacks", where disruptive nodes are selected uniformly at random. In contrast, Sec.\ \ref{Targeted Attacks: Unveiling the Impact of Targeting High-Degree Nodes} considers ``targeted attacks," where the disruptive nodes are selected among the most well connected nodes. %The Secs.\ \ref{Untargeted attacks} and\ \ref{Targeted Attacks: Unveiling the Impact of Targeting High-Degree Nodes} have multiple subsections, each focusing on a specific network topology: (i) Regular \cite{harary2018graph}, (ii) Random \cite{gilbert1959random}, (iii) Small world \cite{watts1998collective}, and (iv) Scale-free \cite{barabasi2009scale}.
Sections \ref{Untargeted attacks} and \ref{Targeted Attacks: Unveiling the Impact of Targeting High-Degree Nodes} comprise various subsections, each dedicated to exploring a distinct network topology, namely: (i) Regular \cite{harary2018graph}, (ii) Random \cite{gilbert1959random}, (iii) Small-world \cite{watts1998collective}, and (iv) Scale-free \cite{barabasi1999emergence,barabasi2009scale}. For each attack strategy and each network topology, we derive the threshold of synchronization analytically and compare it with extensive numerical calculations. In Sec.\ \ref{Targeted Attacks: Unveiling the Impact of Targeting Low-Degree Nodes}, we briefly discuss the scenario of targeting low degree nodes. Next, in Sec.\ \ref{Exploring opinion dynamics: random walk of polarized agents} we move on to the second application, the cusp catastrophe model of opinion dynamics. Section\ \ref{Cusp catastrophe as internal dynamics of mobile agents} introduces the applicable free evolution and interaction functions and computes analytically the condition of consensus formation among mobile agents under untargeted attacks. We additionally confirm the result through numerical validation. Finally, Sec.\ \ref{DISCUSSION} presents the discussion and conclusions.

%To validate our analytical findings, we conduct extensive numerical simulations for synchronization, present 
%
%, and polarized opinions, presented in Sec.\ \ref{Exploring opinion dynamics: random walk of polarized agents}. In these simulations, we explore two different approaches to introduce nodes with repulsive couplings. The first approach involves placing disruptive nodes in the complex networks uniformly at random. Alternatively, we adopt a more sophisticated strategy by targeting higher-degree nodes for this repulsive interaction. 
%
%To validate our findings and provide a comparative overview, we select four distinct network topologies: (i) Regular \cite{harary2018graph}, (ii) Random \cite{gilbert1959random}, (iii) Small world \cite{watts1998collective}, and (iv) Scale-free \cite{barabasi2009scale}. Through the examination of these diverse network structures, we further consolidate the robustness of our results and gain insights into the relative performance of different topologies in the face of disruptive nodes. By delving into analytical derivations, conducting detailed numerical simulations, and leveraging various network topologies, we aim to shed light on the intricate dynamics of coherence among mobile agents. This holistic approach expands our understanding of complex systems and facilitates the identification of key factors that govern the emergence of coherent behavior within such networks.

\section{Mathematical Model} \label{Mathematical Model}

We consider a finite network of $n$ vertices. The connectivity of this network is characterized by an adjacency matrix $A=[A_{\alpha\beta}]_{n \times n}$, where $A_{\alpha\beta}=1$ (or $0$) indicates the presence (or absence) of a link between nodes $\alpha$ and $\beta$. We also impose the following assumptions on the network: it is connected, undirected ($A_{\alpha\beta}=A_{\beta\alpha}$), devoid of self-loops ($A_{\alpha\alpha}=0$). The degree of a node $\alpha$ is given by the conventional way, expressed as $d_{\alpha}=\sum_{\beta=1}^{n}A_{\alpha \beta}$. Greek indices number the nodes, while Latin indices are reserved for enumerating the agents which we discuss next.
  
We randomly place $N$ mobile agents on this network of $n$ nodes. After every fixed time interval $\Delta T$, each agent jumps to one of the nodes adjacent to its current position. Consider the $i$-th agent, located at node $\alpha$ at time $t$. At the time $(t+\Delta T)$, this agent will hop to one of the node-$\alpha$'s neighbor nodes, say $\beta$, with a uniform probability $\frac{A_{\alpha \beta}}{d_{\alpha}}$. Once the agent has made its jump, it interacts with all the other agents present in node $\beta$ at that time. The state $\phi_i$ of the mobile agent $i$ $(i=1,2,3,\cdots,N)$, situated on node $\alpha$ $(\alpha=1,2,3,\cdots,n)$, evolves according to the following equation:

\begin{equation}\label{1}
	\begin{split}
		\dot{\phi}_i=F_i(\phi_i)+ \sum_{j \in O_{\alpha}} H_\alpha(\phi_i,\phi_j).
	\end{split}
\end{equation}
The term $F_i(\phi_i)$ describes the agents natural, free evolution. The subscript of $F_i$ explicitly enumerates the diversity of the agents. $H_\alpha(\phi_i,\phi_j)$ describes how agents interact with each other within node $\alpha$. The subscript of $H_\alpha$ enumerates the diversity of nodes or locations. At every time instance, each node $\alpha$ hosts a particular subset of mobile agents $O_{\alpha}$, and these subsets change after random walk iterations. We continue this hopping process, which involves local interactions, for a significant number of iterations until a stationary state is reached. 

%Thus, our model comprises three different aspects. First, the internal state of each agent evolves depending on their individual first order differential equation. Second, the internal state is influenced by all other agents in the same node, i.e., the local neighborhood, which changes dynamically as the agents move. And third, their collective motion, akin to swarming, is modeled as an unbiased random walk. The interactions take an arbitrary form, and vary from node to node to account for differences between interaction venues.

%By delving into this temporal network model, we aim to illuminate the intricate interplay between mobile agents' motions and their corresponding internal states, filling a significant void in the current research landscape. In the following section, we will use analytical methods to compute the effects of network topology and random walk explicitly in this generic set-up.

\section{Analytical findings} \label{Mean-field analysis}

 To start our analysis, we use a master equation describing the random walk of a single mobile agent on the network. Let us assume that the random walk begins at a node $\delta$. We use the notation $P_{\alpha\delta}(t)$ to represent the probability of finding the agent at node $\alpha$ after a specific time $t$. This probability can be expressed recursively using the master equation

\begin{equation}\label{2}
	\begin{split}
		P_{\alpha \delta}(t)=\sum_{\beta} A_{\alpha \beta} \dfrac{P_{\beta \delta}(t-\Delta T)}{d_{\beta}}.
	\end{split}
\end{equation}

The random walk iterations occur regularly at intervals of $\Delta T$, therefore, $P_{\beta \delta}(t-\Delta T)$ represents the probabilities during the last iteration. The master equation states that the agent will be in node $\alpha$ if, during the previous iteration, it was in one of the neighboring nodes $\beta$ (probability given by $P_{\beta \delta}(t-\Delta T)$), and then it jumped to node $\alpha$ (probability given by $\frac{1}{d_\beta}$). As the time $t$ approaches infinity, Eq.\ \eqref{2} reaches a stationary state where the probability distribution becomes independent of time $t$ and the starting node $\delta$ \cite{noh2004random}. Thus, in the stationary state, we have the following equation,
 \begin{equation}\label{3}
 	\begin{split}
 		P_{\alpha}=\sum_{\beta} A_{\alpha \beta} \dfrac{P_{\beta}}{d_{\beta}}.
 	\end{split}
 \end{equation}

We can easily verify that $\frac{P_{\alpha}}{d_{\alpha}}=c$ (with $c$ independent of $\alpha$) is a solution of Eq.\ \eqref{3}. Pulling out $\frac{P_{\beta}}{d_{\beta}}$ as a common factor, the remaining sum evaluates to $d_{\alpha}$. Therefore, the expression $\frac{P_{\alpha}}{d_{\alpha}}=c$ satisfies the stationary state condition. After normalization, the probability of finding the specific agent in node $\alpha$ can be expressed as

\begin{equation}\label{4}
	P_{\alpha}=\dfrac{d_{\alpha}}{\sum_{\beta} d_{\beta}}.
\end{equation}

 In a connected network, any node can be reached from any other node. This means that the Markov chain corresponding to such a random walk is irreducible and therefore the stationary state given by Eq.\ \eqref{4} must be unique.

Considering that we have a total of $N$ agents, the expected number of agents in node $\alpha$ can be calculated using the following equation:

\begin{equation}\label{5}
	|O_{\alpha}|=\dfrac{N d_{\alpha}}{\sum_{\beta} d_{\beta}}.
\end{equation}

Next, we will utilize the averaging theory to establish the equations for the internal states of the mobile agents in the weak coupling limit. We define the weak coupling limit by demanding that the interaction time-scale is much slower than the random walk time-scale. In this case the averaging theory \cite{sanders2007averaging} allows us to replace the weak, fast-shifting interaction terms in Eq.\ \eqref{1} with their averaged values over all agents. This approximation becomes exact in the limit of infinitely separated time scales, where $\Delta T \to 0+$. One can equivalently interpret this as a rapid random walk instead of weak interactions. In terms of opinion dynamics, this limit is justified by the fact that changing ones opinion significantly is unlikely after just one interaction. By averaging over all possible neighbors, Eq.\ \eqref{1} can be simplified as follows

\begin{equation}\label{6}
\begin{split}
		\dot{\phi}_i&=F_i(\phi_i)+ \sum_{j \in O_{\alpha}} \dfrac{1}{N} \sum_{l=1}^{N} H_\alpha(\phi_i,\phi_l),\\
	&=F_i(\phi_i)+|O_{\alpha}| \dfrac{1}{N} \sum_{l=1}^{N} H_\alpha(\phi_i,\phi_l),\\
	&=F_i(\phi_i)+\dfrac{d_{\alpha}}{\sum_{\beta=1}^{n}d_{\beta}} \sum_{l=1}^{N} H_\alpha(\phi_i,\phi_l).
\end{split}
\end{equation}

Recall that, in the given expression, node $\alpha$ represents the current location of agent $i$. Building upon the reasoning discussed earlier; we can proceed by averaging the interaction terms originating from node $\alpha$ across all possible nodes $\delta$ that the agent could visit. This averaging considers the appropriate probability weights $P_{\delta}$ associated with each node $\delta$. Hence, we obtain

\begin{equation}\label{7}
	\dot{\phi}_i=F_i(\phi_i)+ \sum_{\delta=1}^{n} P_{\delta} \dfrac{d_{\delta}}{\sum_{\beta=1}^{n}d_{\beta}} \sum_{l=1}^{N} H_\delta(\phi_i,\phi_l).
\end{equation}

To make further progress, we assume that the interaction functions of different nodes relate to each other through scaling $H_\delta(\phi_i,\phi_j)=k_\delta H(\phi_i,\phi_j)$, where $k_\delta$ is the coupling strength between the mobile agents in node $\delta$. %The couplings $k_{\delta}$ are randomly sampled from a distribution $f_{\delta}(k)$. Each node can have a different distribution (thus the subscript $\delta$ on $f_\delta(k)$) depending on its characteristics, e.g., its degree. This distribution variability lets us investigate how the number of connections affects the node's ability to influence the global state. Once the coupling strengths $k_{\delta}$ are assigned to nodes at the beginning, they remain constant throughout the entire process.

\begin{equation}\label{8}
	\begin{split}
		\dot{\phi}_i&=F_i(\phi_i)+ \sum_{\delta=1}^{n} P_{\delta} \dfrac{k_{\delta}d_{\delta}}{\sum_{\beta=1}^{n}d_{\beta}} \sum_{l=1}^{N} H(\phi_i,\phi_l),\\
		&=F_i(\phi_i)+   \dfrac{\sum_{\delta=1}^{n} k_{\delta} {d_{\delta}}^2}{\big({\sum_{\beta=1}^{n}d_{\beta}\big)}^2} \sum_{l=1}^{N} H(\phi_i,\phi_l),\\
		&=F_i(\phi_i)+   \dfrac{\frac{1}{n}\sum_{\delta=1}^{n} k_{\delta} {d_{\delta}}^2}{n \big({\frac{1}{n}\sum_{\beta=1}^{n}d_{\beta}\big)}^2} \sum_{l=1}^{N} H(\phi_i,\phi_l),\\
		&=F_i(\phi_i)+ \dfrac{\langle d^2 k \rangle}{n {\langle d \rangle}^2}   \sum_{l=1}^{N} H(\phi_i,\phi_l),\\
		&=F_i(\phi_i)+ \dfrac{\tilde{k}}{N}   \sum_{l=1}^{N} H(\phi_i,\phi_l).
	\end{split}
\end{equation}

In the given expression, the notation $\langle \cdot \rangle$ represents a simple, unweighted average taken over all nodes. It is important to observe that the outcome is a differential equation resembling the original equation \eqref{1}. However, this time the system is globally coupled with an effective coupling strength denoted as $\tilde{k}$. The resulting effective dynamical equation is

\begin{equation}\label{9}
	\begin{split}
		&\dot{\phi}_i=F_i(\phi_i)+ \dfrac{\tilde{k}}{N}   \sum_{j=1}^{N} H(\phi_i,\phi_j) 
		\\
		&\tilde{k}=\dfrac{N}{n} \dfrac{\langle d^2 k \rangle}{ {\langle d \rangle}^2}.
	\end{split}
\end{equation}

%In this equation, $N$ represents the total number of agents, $n$ is the number of nodes, and $\langle d^2 k \rangle$ and $\langle d \rangle^2$ are the averages of the squared degrees and degrees, respectively.
These effective differential equations for agent states are the main analytic result of our findings. The effective coupling strength $\tilde{k}$ depends on several factors including the size of the network, the number of agents, the network topology, and the distribution of the coupling strengths. Therefore, by varying any of these parameters, the effective coupling strength can be altered, leading to different dynamics and collective behaviors in the network of interacting mobile agents. Below, we will discuss the insights readily available from Eq.\ \eqref{9}, as well as its detailed consequences for different applications in Secs.\ \ref{Exploring Synchronization: Random walk of Kuramoto agents} and\ \ref{Exploring opinion dynamics: random walk of polarized agents}.

\par At this stage, several observations can be made without delving into the specific details of the dynamics. 

\begin{itemize}
	\item {\it First, due to the weighting by the squared node degree, nodes with higher degrees have a greater influence on the system's behavior. These highly connected nodes play a more significant role in shaping the overall dynamics of the system.}
	
	\item {\it  Second, when considering the number of agents $N$ and the network size $n$, an inverse relationship can be observed. As the number of agents increases and the network size decreases, the agents become more concentrated within the network. This concentration leads to a higher frequency of interactions among any given pair of agents, resulting in an increased effective coupling strength. This density-dependent synchronization threshold is closely linked to phenomena like bacterial infection, biofilm formation, and bioluminescence, unveiling quorum-sensing transitions in coupled systems \cite{nadell2008evolution,taylor2009dynamical,camilli2006bacterial,chowdhury2019synchronization20}.}
	
	\item {\it Third, the term $\langle d^2 k \rangle$ in the effective coupling informs an intelligent design of the network structure, with nodes' inherent variations in mind. In particular, correlating the node degrees with their coupling strengths enhances the effective interactions.}
	
\end{itemize}

% In summary, the observations highlight the importance of node degrees and the interplay between the number of agents and network size in determining the dynamics and behavior of the system.

 Finally, it is informative to examine the scenario where all nodes possess an identical positive coupling strength. In this case, the coupling term can be factored out of the expectation, resulting in $\langle d^2 k \rangle = k \langle d^2 \rangle$. Consequently, the expression for the effective coupling in Eq.\ \eqref{9} simplifies to:

\begin{equation}\label{10}
	\tilde{k} = \dfrac{N}{n} \dfrac{\langle d^2 \rangle}{\langle d \rangle^2} k.
\end{equation}

 This observation reveals a counter-intuitive finding: degree heterogeneity, which refers to variation in the number of connections among network nodes, actually enhances the effective coupling and promotes coherence among the agents. On the other hand, when network nodes have similar degrees, the similarity in agent states decreases exemplifying the converse symmetry breaking phenomenon \cite{nishikawa2016symmetric}. This result becomes more intuitive when interpreted in the context of opinion dynamics. When there are several highly connected hubs in the network that serve as focal points for discussions or interactions, and most agents are concentrated within these hubs, it becomes easier to reach a consensus or synchronization among the agents. In contrast, if there are many small discussion venues or nodes with equal popularity, the process of achieving consensus becomes more challenging. %In essence, {\it having a heterogeneous distribution of node degrees facilitates the formation of coherence among the agents, while a more uniform distribution hinders this process}. This finding provides insights into the dynamics of opinion formation and highlights the role of network structure in influencing collective behavior.

\section{Exploring Synchronization: Random walk of Kuramoto agents}\label{Exploring Synchronization: Random walk of Kuramoto agents}
% In this section, we briefly introduce the synchronization in connection to swarming and discuss the applicability to sociological experiments.
% KURAMOTO HISTORY
The exploration of synchronization has a fascinating history that dates back to Huygens' classical pendulum experiment \cite{willms2017huygens} and Winfree's pioneering work \cite{winfree1967biological} on coupled oscillators for circadian rhythms. Winfree discovered that synchronization spontaneously emerges when the coupling strength between oscillators exceeds a critical value, resembling a phase transition. Building upon this, Kuramoto \cite{kuramoto1975international} simplified the model and derived an exact analytical solution, sparking widespread interest in the dynamics of coupled oscillators \cite{pikovsky2003synchronization}. Kuramoto's model has been extended to various systems beyond circadian rhythms in recent years. Examples include firing neurons \cite{o2016dynamics}, chorusing frogs \cite{aihara2008mathematical}, and even audiences clapping in perfect unison at concerts \cite{neda2000physics}. The study of Kuramoto oscillators synchronizing has also provided insights into diverse phenomena, such as the behavior of power grids \cite{dorfler2012synchronization}, phase locking in Josephson junction arrays \cite{wiesenfeld1996synchronization}, the feedback between the oscillatory and cascading dynamics \cite{mikaberidze2022sandpile}, and even the unexpected wobbling of London's Millennium Bridge on its opening day \cite{strogatz2005crowd}. Remarkable progress has been achieved in understanding how different network structures influence the synchronization behavior of coupled Kuramoto oscillators \cite{rodrigues2016kuramoto}.

% OSCILLATORS VS SWARMING
It is worth highlighting that the domains of swarming and synchronization share numerous commonalities, residing at the intersection of nonlinear dynamics and statistical physics. However, it is regrettable that these fields have remained largely disconnected, calling for additional research attention. In the study of swarming, the primary emphasis lies in understanding how individuals move collectively, often overlooking the internal dynamics within each agent. Conversely, studies on synchronization predominantly delve into the intricacies of oscillators' internal dynamics, paying less attention to their motion. This disparity in focus presents an intriguing opportunity for further exploration and integration of ideas from both fields. By bridging this gap and combining insights from swarming and synchronization \cite{sar2022swarmalators,o2017oscillators}, we can gain a deeper understanding of collective behaviors in complex systems.
 
Below we discuss swarming and synchronizability of oscillators motivated by recent neuro-sociological studies \cite{wheatley2019beyond, hu2017brain, perez2017brain, sievers2020consensus, wohltjen2021eye}. These experiments show that brain activities of interacting individuals get synchronized. We will consider mobile agents that synchronize upon interactions with others. After interacting for some time, they move to other locations in a network of interaction venues, where they interact with a new set of agents, and so on. We represent the agents' internal dynamics and their interactions through the most widely studied model of synchronization, the Kuramoto dynamics.

First, we will describe the specific setup of Kuramoto oscillators in our model (Sec.\ref{Kuramoto model as internal dynamics of mobile agents}). Then, we will compute explicitly the synchronizability condition for various network topologies and compare them with simulations for various attack strategies (Secs.\ \ref{Untargeted attacks}, \ref{Targeted Attacks: Unveiling the Impact of Targeting High-Degree Nodes}, \ref{Targeted Attacks: Unveiling the Impact of Targeting Low-Degree Nodes}).

\subsection{Kuramoto model as internal dynamics of mobile agents}
\label{Kuramoto model as internal dynamics of mobile agents}
 To progress with the analysis, we fix the free evolution function $F_i(\phi_i)=\omega_i$ and the interaction function $H(\phi_i, \phi_j) = \sin(\phi_j - \phi_i)$ in accordance with Kuramoto dynamics. Here, $\omega_i$ represents the natural frequency of agent $i$ sampled from a unimodal, symmetric distribution denoted $g(\omega)$.
We can numerically investigate the system's dynamics and validate our theoretical analysis. 
A video showcasing the random movement and synchronization of such Kuramoto agents can be found at \cite{web_64}. With these choices, Eq.\ \eqref{9} can be expressed as
 
 \begin{equation}\label{11}
 	\begin{split}
 		\dot{\phi}_i=\omega_i+ \dfrac{\tilde{k}}{N}   \sum_{j=1}^{N} \sin(\phi_j-\phi_i).
 	\end{split}
 \end{equation}

To analyze synchronization, we employ the conventional Kuramoto order parameter $r=|\frac{1}{N}\sum_{j=1}^N \exp{(\hat i \phi_j)}|$ where $\hat i=\sqrt{-1}$. Here averaging happens over all $N$ agents. For the incoherent states, the order parameter vanishes in the thermodynamic limit of agents $N\rightarrow\infty$, while once synchronization emerges, order parameter attains a positive value in this limit. The synchronizability condition for the globally coupled Kuramoto oscillators, as described in Ref.\ \cite{rodrigues2016kuramoto}, can be expressed as

\begin{equation}\label{12}
	\tilde{k} > \dfrac{2}{\pi {\lVert g(\omega) \rVert}_\infty}.
\end{equation}

The notation $\lVert g(\omega) \rVert_\infty$ refers to the L-infinity norm of the distribution $g(\omega)$ and is calculated as the maximum value of $g(\omega)$. Combining Eqs.\ \eqref{9} and \eqref{12}, we get the synchronization condition for the full model

\begin{equation}\label{13}
 \dfrac{\langle d^2 k \rangle}{ {\langle d \rangle}^2} > 	\dfrac{n}{N} \dfrac{2}{\pi {\lVert g(\omega) \rVert}_\infty}.
\end{equation}
It depends on the joint degree and coupling distributions through the term $\langle d^2 k \rangle$. The computation of the expectation terms in this equation relies on the specific distributions used to generate the network under consideration. In order to obtain accurate results using this expression, it is necessary to consider the thermodynamic limit of the network size $n\rightarrow\infty$. In this limit, the degree distributions become exact and more accurately represent the statistical properties of the network structure.

 To validate our results through simulations, we need to specify the frequency distribution $g(\omega)$. As is customary in many studies, we select the normal distribution:

\begin{equation}\label{14}
	g(\omega) = \frac{1}{\Delta \omega \sqrt{2\pi}} \exp\left(-\frac{1}{2}\left(\frac{\omega-\omega_0}{\Delta \omega}\right)^2\right).
\end{equation}
Here, $\omega_0$ represents the mean or central value of the frequencies, and $\Delta \omega$ is the standard deviation or width of the distribution. The normal distribution is a widely used choice in modeling various systems, including Kuramoto oscillators, due to its abundance in real world, mathematical tractability, and symmetry.

The maximum value of the normal distribution $g(\omega)$ occurs at the mean frequency $\omega = \omega_0$. By substituting $\omega = \omega_0$ into Eq.\ \eqref{14}, we obtain

\begin{equation}\label{15}
	\lVert g(\omega) \rVert_\infty = g(\omega_0) = \frac{1}{\Delta \omega \sqrt{2\pi}}.
\end{equation}

Consequently, the synchrony condition given in Eq.\ \eqref{13} can be rewritten as

\begin{equation}\label{16}
	\frac{\langle d^2 k \rangle}{{\langle d \rangle}^2} > \sqrt{\frac{8}{\pi}} \frac{n}{N} \Delta \omega.
\end{equation}

In this form, the condition relates the joint degree and coupling distributions to the network size $n$, the number of agents $N$, and the width of the frequency distribution $\Delta \omega$. It provides a criterion for synchronization based on these parameters, indicating the necessary condition for achieving synchronization in the system of Kuramoto oscillators.

% In our analysis, we will examine the implications of Eq.\ \eqref{16} for different network topologies and attack strategies. We begin by considering ``untargeted attack" cases, where nodes are chosen uniformly at random to be corrupted meaning their coupling constant is set to negative (i.e., repulsive). This means that all nodes have equal chance of being assigned repulsive coupling strength. We will investigate the performance of networks with different topologies under this attack strategy and compare our theoretical predictions with numerical simulations.
%
% Next, we will focus on ``targeted attack" scenarios, where the most well connected nodes are corrupted and their couplings set to negative. This attack strategy aims to disrupt synchronization by targeting highly influential nodes. We will analyze the effects of targeted attacks on networks with various topologies and again compare our theoretical findings with numerical simulations to validate our results.
%
% By examining these different attack strategies and network topologies, we aim to gain insights into the system's behavior and understand how synchronization is affected in each case. The comparison between theoretical predictions and numerical simulations will help us verify the accuracy of our theoretical framework and provide a comprehensive understanding of the dynamics in these scenarios.

\subsection{Untargeted attacks}\label{Untargeted attacks}

 We begin our analysis by examining the simplest scenario of untargeted attacks where nodes are chosen uniformly at random to be corrupted, i.e., assigned a negative, repulsive coupling. Then the coupling distribution is not influenced by the node degree, and any node has an equal probability of being repulsive, regardless of its degree. In this case, we can observe that the joint distribution term becomes independent and separates into two individual terms

\begin{equation}\label{17}
	\langle d^2 k \rangle = \langle d^2 \rangle \langle k \rangle.
\end{equation}

This equation simplifies the analysis, allowing us to examine the behavior of each term independently. Consequently, Eq.\ \eqref{16} becomes 

\begin{equation}\label{18}
	\langle k \rangle > \sqrt{\frac{8}{\pi}} \frac{n}{N} \frac{{\langle d \rangle}^2}{\langle d^2 \rangle} \Delta \omega.
\end{equation}

The quantity $\frac{{\langle d \rangle}^2}{\langle d^2 \rangle}$ is always non-negative and it can not exceed $1$ since the nonnegativity of the variance implies  ${\langle d \rangle}^2 \leq \langle d^2 \rangle$.  The extreme cases for this quantity are observed in two types of networks. In regular networks, where each node has the same degree, we have $\frac{{\langle d \rangle}^2}{\langle d^2 \rangle}=1$. On the other hand, in scale-free networks with a degree distribution characterized by a power-law exponent $1<\gamma\leq3$, we find that $\frac{{\langle d \rangle}^2}{\langle d^2 \rangle}=0$ (explicitly derived later in \eqref{scale-free}). {\it In summary, the scale-free networks with this range of power-law exponents are the most robust to untargeted attacks, while regular networks are weakest to untargeted attacks}. This finding perfectly aligns with the structural robustness of the giant connected component in complex networks under random removal of nodes
%. Previous studies on complex network robustness have observed the robustness of scale-free networks with $1<\gamma\leq3$ and the vulnerability of regular networks 
\cite{newman2018networks,molloy1995critical}. By understanding the behavior of the quantity $\frac{{\langle d \rangle}^2}{\langle d^2 \rangle}$ and its implications for network robustness, we gain valuable insights into the interplay between network structure and the impact of untargeted attacks.

 In order to examine how the repulsive couplings impact the system, let us consider the Bernoulli distribution for the coupling strengths. Each node will be assigned a negative coupling strength $k_-$ with probability $p$ (called disruptors or corrupted nodes), or a positive coupling strength $k_+$ with probability $(1-p)$.
 %probability labeled $p$ that describes the likelihood of each node becoming corrupted. When a node becomes corrupted, it obtains a negative coupling strength denoted as ``$k_{-}$". Conversely, the nodes that remain unaffected by corruption possess a positive coupling strength indicated as ``$k_{+}$". 

\begin{equation}\label{19}
	\begin{split}
		&Pr(k=k_{-})=p,\\
		&Pr(k=k_{+})=1-p.
	\end{split}
\end{equation}

Hence, the average coupling becomes

\begin{equation}\label{20}
	\langle k \rangle=p(k_{-}-k_{+})+k_{+}.
\end{equation}

By substituting the value of $\langle k \rangle$ into Eq.\ \eqref{18}, we obtain the expression for the critical fraction $p_c$ of corrupted nodes needed to achieve complete incoherence:

\begin{equation}\label{21}
	p_c=\dfrac{1}{k_{+}-k_{-}} \left(k_{+}-\sqrt{\frac{8}{\pi}} \frac{n}{N} \frac{{\langle d \rangle}^2}{\langle d^2 \rangle} \Delta \omega\right).
\end{equation}

 It should be noted that the equation may yield non-physical values such as $p_c < 0$ or $p_c > 1$. This implies that no critical value of $p_c$ separates the synchronized and incoherent phases. For instance, such a scenario can arise when the positive coupling $k_+$ lacks sufficient strength to synchronize the system, even in the absence of corrupted nodes. Another extreme scenario can occur if the disruptor coupling $k_-$ is set to a positive value $k_->0$, accompanied by a narrow frequency distribution $\Delta\omega \to 0$. This inevitably leads to synchrony.

 Next, we take a closer look at how different network topologies affect the synchronization of mobile agents. We specifically focus on understanding how the critical fraction $p_c$ changes when we use various network structures.

\subsubsection{Regular networks}

\begin{figure}[t]
	\centerline{\includegraphics[scale=0.5]{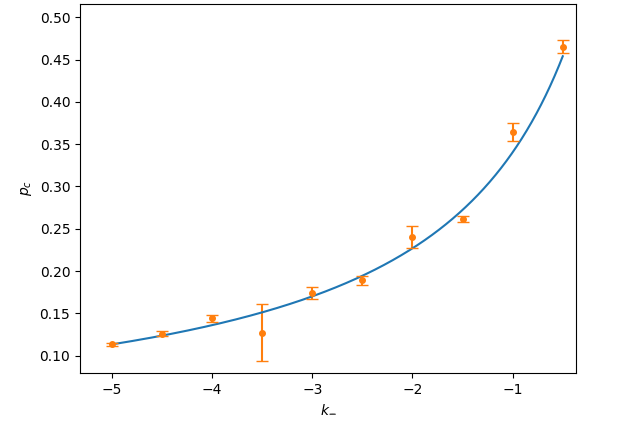}}
	\caption{{\bf Synchronizability of regular networks under untargeted attacks}: As the repulsive coupling strength $k_-<0$ decreases, a smaller fraction $p_c$ becomes sufficient to disrupt the synchronization among the Kuramoto oscillators. The solid curve represents our analytically derived result (Eq.\ \eqref{22}), matching with the numerical simulations in orange data points. Throughout our analysis, we maintain fixed values for the other parameters: $k_+=1$, $d=3$, $\Delta \omega=1$, $n=100$, $N=500$, and $\Delta T=0.001$. To validate our findings, we conduct multiple numerical simulations and plot the results, showing the mean value along with the standard error. This comprehensive approach ensures the robustness and reliability of our conclusions.
	}\label{fig1}
\end{figure}

 For regular networks such as random regular networks, complete graphs, regular lattices, or any other regular network where each node has the same degree $d$, a simplification can be made. In such cases, the expression $\frac{{\langle d \rangle}^2}{\langle d^2 \rangle}$ evaluates to $1$, resulting in the critical coupling equation \eqref{21} being reduced to

\begin{equation}\label{22}
	p_c=\dfrac{1}{k_{+}-k_{-}} \left(k_{+}-\sqrt{\frac{8}{\pi}} \frac{n}{N} \ \Delta \omega\right).
\end{equation}

\begin{figure}[t]
	\centerline{\includegraphics[scale=0.5]{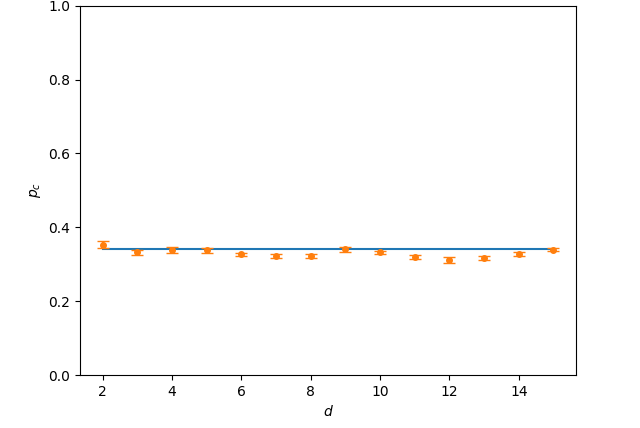}}
	\caption{{\bf Degree independence of synchronizability in regular networks}: The disruption of coherence among the Kuramoto oscillators does not appear to depend on the degree $d$ of each node in the regular network. This observation is similar to our analytical result (Eq.\ \eqref{22}), represented by the solid line in the graph. We maintain a consistent set of parameter values throughout our analysis, with $k_+=1$, $k_-=-1$, $\Delta \omega=1$, $n=100$, $N=500$, and $\Delta T=0.001$.
	}\label{fig2}
\end{figure}

We put our findings to test by comparing them to simulations using regular networks. The outcome is depicted in Fig. \ref{fig1}. As anticipated, when the corrupted nodes possess strong negative couplings, fewer of them are required to create disorder. This figure is drawn by keeping fixed the parameters at $k_+=1$, $d=3$, $\Delta \omega=1$, $n=100$, $N=500$, and $\Delta T=0.001$. Moving forward, Fig. \ref{fig2} reveals that {\it the condition for synchronization remains unchanged, regardless of the network connectivity}. We will see below that what matters instead is the degree fluctuations. Our analytical finding in Eq.\ \eqref{22} aligns well with these numerical simulations, as $p_c$ does not depend on the degree of the regular network. In other words, the ability for synchronization to occur is not influenced by the specific way the network is structured as long as it has a regular degree distribution. The figure was generated using the following parameter values: $k_+=1$, $k_-=-1$, $\Delta \omega=1$, $n=100$, $N=500$, and $\Delta T=0.001$.% These findings shed light on the intricate dynamics within networks and highlight how certain properties of the network, such as connectivity, can affect the behavior of corrupted elements. 

 Throughout our study, the plots serve as visual representations of crucial parameter values that define the boundary between synchronized and incoherent phases. To generate each point on these plots, we keep all parameters fixed except the one represented on the $y$-axis. We then conduct simulations from the beginning, including the generation of the network, for different values of the $y$-axis parameter. During these simulations, we measure the global synchronization order parameter, denoted as $r$, in the stationary state. We observe how $r$ changes as a function of the $y$-axis parameter.

 In the incoherent phase, where the system lacks synchronization, $r$ is equal to $0$. On the other hand, when the system exhibits synchronization, $r$ takes on values greater than $0$, with full synchrony approaching a value close to $1$. With this information, we fit the measured data using a heuristic curve that captures the behavior of $r$. When the incoherent phase lies below the critical point $(y < y_c)$, we employ the curve expression $r(y)=\frac{2}{\pi} \tan^{-1}(c(y-y_c))\theta(y-y_c)$, where $\theta(\cdot)$ represents the Heaviside step function. Conversely, when the incoherent phase is above the critical point $(y > y_c)$, we use the expression $r(y)=\frac{2}{\pi}\tan^{-1}(c(y_c-y))\theta(y_c-y)$.

 The values of $c$ and $y_c$ are determined by fitting the curve to the data using the root-mean-square method. The extracted value of $y_c$ corresponds to the critical value of the $y$-axis parameter. To ensure accuracy, we repeat this entire process multiple times, generating a sample of $y_c$ measurements. In the plots, we present the mean value of this sample, along with the standard error, providing an indication of the reliability and precision of our findings.

\subsubsection{Random networks} 

 The case of random networks is different from regular networks because the degrees of nodes are no longer equal. In fact, the probability that a node has a certain degree can be described by the binomial distribution \cite{newman2018networks}

\begin{equation}\label{23}
	Pr(d)= {n-1 \choose d} \kappa^d (1-\kappa)^{n-1-d}.
\end{equation}
Here $n$ represents the size of the network and $\kappa$ is the probability that two randomly chosen nodes are connected.

We can calculate the average degree $\langle d \rangle$ and average squared degree $\langle d^2 \rangle$ of the network using the following formulas:

\begin{equation}\label{24}
	\begin{split}
		\langle d \rangle&=\kappa(n-1),\\
		\langle d^2 \rangle&=\kappa(1-\kappa)(n-1)+\kappa^2(n-1)^2.
	\end{split}
\end{equation}
With this, the synchronization condition \eqref{21} reduces to

\begin{equation}\label{25}
	p_c=\dfrac{1}{k_{+}-k_{-}} \left(k_{+}-\sqrt{\frac{8}{\pi}} \frac{n}{N} \frac{\kappa(n-1)}{1-\kappa+\kappa(n-1)} \Delta \omega\right).
\end{equation}

\begin{figure}[t]
	\centerline{\includegraphics[scale=0.5]{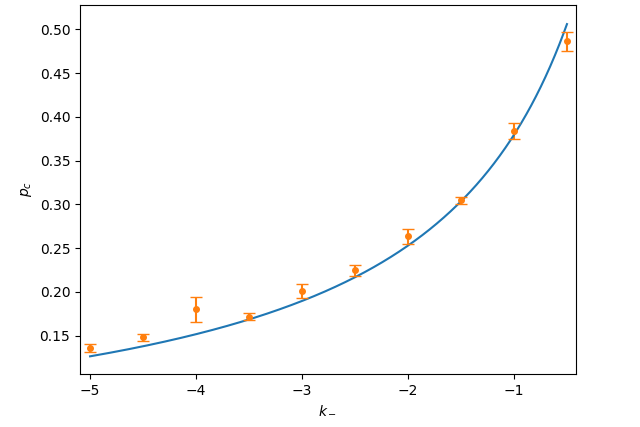}}
	\caption{{\bf Synchronizability of random networks under untargeted attacks}: Critical fraction of disruptors necessary to destroy synchrony as a function of disruptors' coupling strength on a random network. Solid curve presents our analytical result Eq.\ \eqref{25} while the datapoints come from numerical simulations. Fixed parameter values are $k_+=1$, $\kappa=0.03$, $\Delta \omega=1$, $n=100$, $N=500$, and $\Delta T=0.001$.%By randomly corrupting fractions $p_c$ of nodes in a random network, it becomes effortless to disrupt the phase coherence among the mobile agents. Remarkably, our analytical result (Eq.\ \eqref{25}) aligns perfectly with the numerical findings. Notably, an inverse relationship between $p_c$ and $k_{-}$ emerges, indicating that higher values of the repulsive coupling strength $k_{-}$ require a smaller fraction $p_c$ to disrupt the coherence among the Kuramoto oscillators. Throughout our analysis, we maintain a consistent set of parameters: $k_+=1$, $\kappa=0.03$, $\Delta \omega=1$, $n=100$, $N=500$, and $\Delta T=0.001$.
	}\label{fig3}
\end{figure}

 Interestingly, when the condition $\kappa \neq 1$ is satisfied, we observe that the ratio $\frac{{\langle d \rangle}^2}{\langle d^2 \rangle}=\frac{\kappa(n-1)}{1-\kappa+\kappa(n-1)}$ is strictly less than $1$. This finding has significant implications; it indicates that the critical probability $p_c$ is higher for random networks compared to regular networks: {\it random networks generally exhibit greater robustness and are capable of withstanding a larger number of corrupted nodes than regular networks}. It is worth noting that when $\kappa = 1$, random networks become fully connected and therefore regular, hence both expressions \eqref{22} and \eqref{25} yield the same results. To provide visual evidence supporting this observation, we have included a comparison with simulations in Fig. \ref{fig3} where we keep fixed the parameter values at $k_+=1$, $\kappa=0.03$, $\Delta \omega=1$, $n=100$, $N=500$, and $\Delta T=0.001$. The close correspondence between the analytical predictions and the numerical data further validates the accuracy and reliability of our theoretical framework. %Even though some nodes have repulsive coupling strength $k_-<0$, it is still possible to observe phase coherence among mobile agents by selecting a value of $p_c$ below the critical threshold (the solid line in Fig. \ref{fig3}). This synchronization emerges due to occasional interactions between the mobile agents, which vary over time. This dynamic nature of interactions plays a crucial role in enabling synchronization, which is a challenging feat to achieve in static networks with attractive-repulsive interactions.

\subsubsection{Small world networks}

\begin{figure}[t]
	\centerline{\includegraphics[scale=0.5]{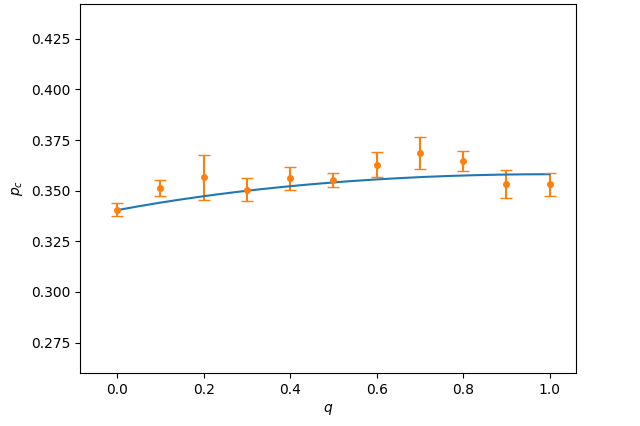}}
	\caption{{\bf Synchronizability of small world networks under untargeted attacks}: We begin with a lattice structure where each node has a degree of $2K$. We rewire the links with a probability $q$.  While keeping other parameters fixed at $n=100$, $K=2$, $k_-=-1$, $k_+=1$, $N=500$, $\Delta \omega=1$, and $\Delta T=0.001$, we generate various small world networks by varying the value of $q$. Subsequently, we plot the numerically simulated $p_c$ for each of these networks alongside the analytical findings (Eqs.\ \eqref{21} and \eqref{26}). Remarkably, the results from our numerical simulations exhibit an impressive agreement with our theoretical analysis. This confirms the accuracy and reliability of our analytical predictions.
	}\label{fig4}
\end{figure}

 In the case of the Watts-Strogatz model for small-world networks, the degree distribution \cite{barrat2000properties} is described by the equation:

\begin{equation} \label{26}
	\begin{split}
		\Pr(d) =& e^{-qK}\sum_{m=0}^{\min(d-K, K)}  {K \choose m} (1-q)^m q^{K-m}
		\\
		&\times \frac{(Kq)^{d-K-m}}{(d-K-m)!} ,\quad\text{for}\quad d\ge K
	\end{split}
\end{equation}
where $2K$ represents the degree of the original lattice (before rewiring), and $q$ is the rewiring probability. We could not obtain a closed form expression for the synchronization threshold in this case due to the complicated nature of the degree distribution. Instead, we utilize Eq.\ \eqref{26} to numerically determine the expectations $\langle d \rangle$ and $\langle d^2 \rangle$ and subsequently apply them in Eq.\ \eqref{21} to predict the critical fraction of corrupted nodes. The results, complemented by simulation data, are illustrated in Fig. \ref{fig4} and show an agreement. The simulations in Fig. \ref{fig4} are produced by maintaining set parameter values: $n=100$, $K=2$, $k_-=-1$, $k_+=1$, $N=500$, $\Delta \omega=1$, and $\Delta T=0.001$.

\subsubsection{Scale-free networks} \label{scale-free}

 When we look at scale-free networks, where the node degree follows a power-law distribution $Pr(d) \propto d^{- \gamma}$, we discover that the system behavior changes qualitatively depending on the value of the exponent $\gamma$. For $2<\gamma \leq 3$, the second moment $\langle d^2 \rangle$ diverges while the first moment $\langle d \rangle$  is finite. And hence, the ratio ${\langle d^2 \rangle}/{{\langle d \rangle }^2}$ vanishes. Similarly, for $1<\gamma \leq 2$ we have $\frac{\langle d^2 \rangle}{{\langle d \rangle }^2} \approx \frac{(2-\gamma)^2}{(\gamma-3)(\gamma-1)} \lim\limits_{D \to \infty} \frac{(D^{2-\gamma}-1)^2}{D^{3-\gamma}-1}=0$. Thus, Eq.\ \eqref{21} reduces to

\begin{equation}\label{27}
	p_c=\dfrac{k_+}{k_{+}-k_{-}}.
\end{equation}
However, for $\gamma>3$, we have

\begin{equation}\label{28}
	\begin{split}
		\langle d \rangle&=\dfrac{\zeta(\gamma-1)}{\zeta(\gamma)},\\
		\langle d^2 \rangle&=\dfrac{\zeta(\gamma-2)}{\zeta(\gamma)},
	\end{split}
\end{equation}
where $\zeta(\gamma)$ is the Riemann zeta function. Then, Eq.\ \eqref{21} yields

\begin{equation}\label{29}
	p_c=\dfrac{1}{k_{+}-k_{-}} \left(k_{+}-\sqrt{\frac{8}{\pi}} \frac{n}{N} \frac{{\zeta(\gamma-1)}^2}{\zeta(\gamma-2)\zeta(\gamma)} \Delta \omega\right).
\end{equation}

\begin{figure}[t]
	\centerline{\includegraphics[scale=0.5]{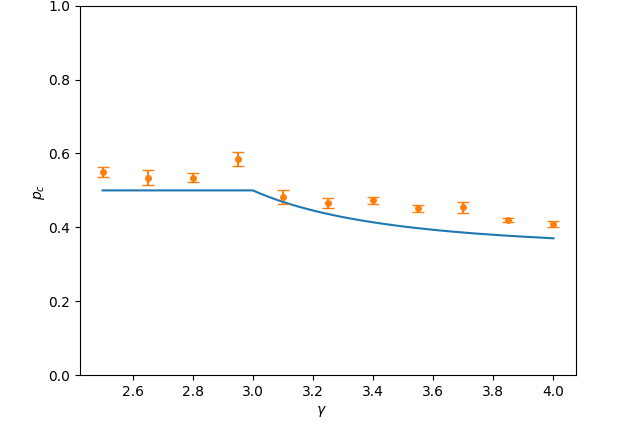}}
	\caption{{\bf Synchronizability of scale-free networks under untargeted attacks}: Our analytical result (Eq.\ \eqref{27}) reveals that the critical fraction $p_c$ of nodes with repulsive coupling, beyond which synchronization becomes unattainable, remains constant regardless of the power law degree exponent $\gamma$, as long as $1 < \gamma \leq 3$. As a result, when $\gamma$ falls within the range of $(1,3]$, we observe a horizontal line in our analysis, while for $\gamma > 3$, the solid line (Eq.\ \eqref{29}) demonstrates a decreasing trend. We conduct numerical simulations on scale-free networks with $n=1000$ vertices and $N=5000$ mobile agents to further verify this analytical understanding. %These simulations serve to validate and strengthen our analytical insights regarding the independence of the synchronization critical fraction on the power law degree exponent.
	Other parameters are kept fixed at $k_-=-1$, $k_+=1$, $\Delta \omega=1$, and $\Delta T=0.0001$. 
	}\label{fig5}
\end{figure}

Figure \ref{fig5} presents numerical data on how $p_c$ depends on $\gamma$. For values of $\gamma$ below 3, $p_c$ remains constant. Above 3, agents become easier to desynchronize, resulting in a lower value of $p_c$. Even though the analytical curve and the numerical data show the same trend, the curve is clearly outside the error bars. This happens because the error bars show the precision of the numerical data and not the accuracy. The accuracy, on the other hand, is controlled by the extent to which we were able to reproduce the infinite time-scale separation and the thermodynamic limits of network size and agent numbers. For the thermodynamic limits, one would need to send $N\rightarrow \infty$ and $n\rightarrow \infty$, keeping $n/N=\it{constant}$ all the while. And the infinite time-scale separation is attained by sending $\Delta T\rightarrow 0$. Improving the simulations in either of these aspects is computationally costly and can be realized only to an extent. This figure was created using specific parameters: $n=1000$, $k_-=-1$, $k_+=1$, $N=5000$, $\Delta \omega=1$, and $\Delta T=0.0001$.  In other words, compared with previous examples, we increased $n$ and $N$, and decreased $\Delta T$ by one order of magnitude each. The numerical results gradually approach our theoretical findings. Yet, we still see the finite size effects in Fig. \ref{fig5}. This should not be a surprise since scale free networks are extremely sensitive to finite size effects \cite{serafino2021true,boguna2004cut}. This topic will be discussed further in Sec.\ \ref{Scale-free networks_targeted}.

%However, as we increase the values of $N$ and $n$, and decrease $\Delta T$, the numerical simulations on scale-free networks gradually approach our theoretical findings. To overcome any limitations due to finite network size, we require large values of $N$ and $n$, as well as small $\Delta T$. Our analysis holds true when the time interval $\Delta T$ tends to zero (indicating rapid movement of mobile agents) and in the thermodynamic limit where $n$ approaches infinity in an extremely large network. Therefore, to achieve accurate results that match our theoretical predictions, we need to use large values of $N$ and $n$, while minimizing $\Delta T$.

\subsection{Targeted Attacks: Unveiling the Impact of Targeting High-Degree Nodes} \label{Targeted Attacks: Unveiling the Impact of Targeting High-Degree Nodes}

 In this new approach, we aim to strategically assign the repulsive coupling strength $k_-$ by targeting the highest degree nodes in the networks. These nodes are particularly influential as they have a greater impact on the collective behavior. To implement this strategy, we sort the nodes in ascending order based on their degrees and select a fraction $p$ from the end of this sorted list. The nodes in this selected fraction will be assigned a negative coupling $k_-$, while the remaining nodes will have a positive coupling $k_+$. This targeted assignment ensures that the most highly connected nodes, which have the potential to disrupt synchronization more efficiently, are equipped with the negative coupling, while other nodes maintain a positive coupling.

 To determine the synchronization condition in this targeted attack scenario, we must calculate the term $\langle d^2 k\rangle$ that appeared in Eq.\ \eqref{13}. First, we determine the cutoff degree $d_c \in \mathbb{Z}$ beyond which nodes are targeted and assigned with negative coupling $k_-$. This cutoff is determined by the consistency equation:

\begin{equation}\label{30}
	p = \sum_{d=d_c}^\infty \Pr(d) = 1 - \Pr(d\leq d_c),
\end{equation}
where $\Pr(d\leq d_c)$ represents the cumulative probability distribution of node degrees. It's important to note that $p$ represents a fraction of nodes with values ranging from $0$ to $1$, while $d$ denotes integer values for node degrees. Consequently, there might not be a clean integer cutoff $d_c$ that isolates an arbitrary fraction $p$ of all nodes. In such cases, we can resort to continuous approximations of the sums or, when applicable, start the summation at $d=\lceil d_c \rceil$ and include only the fraction $\lceil d_c \rceil - d_c$ of nodes with degree $\lfloor d_c \rfloor$. Similar interpretations apply to sums with non-integer bounds.

For the targeted attack scenario, the expression for the joint distribution term becomes:

\begin{equation}\label{31}
	\begin{split}
		\langle d^2 k \rangle = k_+ \sum_{d=1}^{d_c-1} d^2 \Pr(d) + k_- \sum_{d=d_c}^\infty d^2 \Pr(d).
	\end{split}
\end{equation}

 By utilizing this expression along with Eq.\ \eqref{16}, we can determine the critical coupling strength $k_-^c$ required to disrupt synchronization for the corrupted nodes:

\begin{equation}\label{32}
	k_-^c = \frac{-1}{\overset{\infty}{\underset{d=d_c}{\sum}} d^2 \Pr(d)} \left(k_+ \sum_{d=1}^{d_c-1} d^2 \Pr(d) - \frac{n\sqrt{8}\langle d \rangle^2 \Delta \omega}{N\sqrt{\pi}} \right).
\end{equation}

 %Comparing targeted attacks to untargeted attacks, we can demonstrate that the former is at least as effective as the latter.
 The advantage of targeting higher-degree nodes becomes evident when comparing the term $\frac{\langle d^2 k \rangle}{\langle d \rangle^2}$ (the left-hand side of Eq.\ \eqref{13}) for the two attack strategies. By sorting the nodes $\alpha$ based on their degrees (non-decreasing order), we observe that since the highest degree nodes possess the negative coupling $k_-$, the sequence $k_\alpha$ becomes non-increasing. Utilizing Chebyshev's sum inequality, we obtain:

\begin{equation}\label{33}
	\begin{split}
		\frac{\langle d^2 k \rangle}{\langle d \rangle^2} \leq \frac{\langle d^2 \rangle \langle k \rangle}{\langle d \rangle^2}.
	\end{split}
	\end{equation}
	
We recognize the upper bound in Eq.\ \eqref{33} as the left-hand side of Eq.\ \eqref{13} for the untargeted attack scenario. This inequality indicates that {\it the synchronization condition \eqref{13} is more difficult to satisfy under targeted attacks, signifying a weaker network robustness in this case}. It is worth noting that the inequality in Eq.\ \eqref{33} is not strict, and some networks may exhibit equal robustness against both types of attacks.

\subsubsection{Regular networks}

In the case of regular networks, degree-targeted attacks do not provide any advantage over untargeted attacks. This happens due to the absence of any strategic targets in regular networks, where each node contributes equally to the global dynamics. Thus the synchronization condition remains governed by Eq.\ \eqref{22}.

As we found in the last section, the system is less robust under targeted attacks than under untargeted attacks. The more heterogeneous the degrees, the more strategic targets exist. Thus regular networks represent an edge case with no added benefit from targeting, and as we will see later, scale-free networks with $1<\gamma\le 3$ become the least robust under targeted attacks. All this may suggest that regular networks should be the most robust topologies under targeted attack, but this is not so.
%regular networks being the weakest networks under untargeted attacks and their robustness is identical under both attack strategies, it does not necessarily make them the most robust topologies under targeted attacks. 
The reason for this lies in the interplay between degree heterogeneity and the synchronization ability of nodes with positive couplings. While degree heterogeneity enhances the influence of highly connected corrupted nodes with repulsive coupling strength $k_-<0$, it also improves the synchronization capability of nodes with positive couplings. As a result, the overall effect is not straightforward. Even under targeted attacks, heterogeneous networks may exhibit easier synchronization compared to regular networks. The degree distribution of the most robust network depends on various factors, such as the fraction $p$ of corrupted nodes, the coupling strengths, and the distribution of frequencies. The dynamics of the system play a crucial role in determining the specific characteristics of the most robust network structure.

\subsubsection{Random networks} 

\begin{figure}[t]
	\centerline{\includegraphics[scale=0.5]{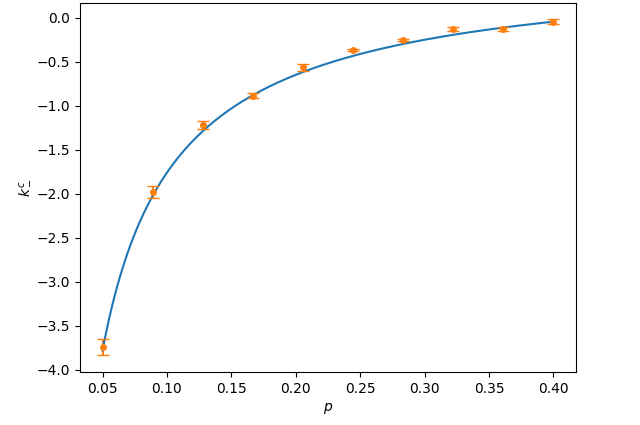}}
	\caption{{\bf Synchronizability of random networks under targeted attacks}: In this figure, we set the parameters as follows: $n=100$, $k_+=1$, $\Delta \omega=1$, $\kappa=0.05$, $N=500$, and $\Delta T=0.001$. The solid line represents our analytical result (see Eqs.\ \eqref{32} and \eqref{35}), which is well aligned with the numerical simulations. %The plot clearly illustrates that as the fraction $p$ of disruptors decreases the critical value of the repulsive coupling strength $k_-^c$ increases. This means that a smaller fraction of nodes is required to disrupt the synchronization among the mobile agents when the negative coupling strength is higher. %The analytical result, represented by the solid line, perfectly matches the numerical simulations, validating the theoretical understanding of the system. This inverse relationship between $k_-^c$ and $p$ highlights the impact of the repulsive coupling strength on synchronization in the presence of targeted attacks. Higher values of $k_-$ make the system more susceptible to disruption, requiring a smaller fraction of nodes to be corrupted.
	}\label{fig6}
\end{figure}

 In random networks, the degrees of nodes are distributed binomially, as described by Eq.\ \eqref{23}. Analytically working with the binomial distribution can be challenging, so we make use of the normal approximation $\Pr(d)\approx \frac{1}{\sigma\sqrt{2\pi}}\exp\left(-\frac{(x-\mu)^2}{2\sigma^2}\right)$. The mean of this approximation is given by $\mu=(n-1)\kappa=\langle d\rangle$, and the standard deviation is $\sigma=\sqrt{(n-1)(1-\kappa)\kappa}$. With this approximation, Eq.\ \eqref{30} can be expressed as:

\begin{equation} \label{34}
	\begin{split}
	p& =1-\Pr(d\le d_c) \\
	&=	1-\frac{1}{2}\left( 1 + \erf\left(\frac{d_c-\mu}{\sigma\sqrt 2}\right)\right)\\ 
	&= \frac{1}{2}\left( 1 - \erf\left(\frac{d_c-\mu}{\sigma\sqrt 2}\right)\right).
	\end{split}
\end{equation}
Solving for $d_c$, we obtain:

\begin{equation} \label{35}
	\begin{split}
		d_c 
		&=
		\mu+\sigma\sqrt 2\erf^{-1}(1-2p)
		\\ &=
		(n-1)\kappa+\sqrt{2(n-1)(1-\kappa)\kappa}\erf^{-1}(1-2p).
	\end{split}
\end{equation}

 Using these expressions, we can compute the critical corrupted coupling strength $k_-^c$ through Eq.\ \eqref{32}. However, the closed-form solution, obtained by evaluating the sums as integrals of the normal approximation, is lengthy and not explicitly presented here. Figure\ \ref{fig6} provides a comparison between the analytical results and simulation data. In this figure, we set the parameters as follows: $n=100$, $k_+=1$, $\Delta \omega=1$, $\kappa=0.05$, $N=500$, and $\Delta T=0.001$. The plot shows that as the magnitude of $k_-^c$ increases, the critical fraction $p$ decreases. This means that only a smaller fraction of nodes with a higher repulsive coupling strength is needed to disrupt the synchronization among the mobile agents. The trend observed in this figure is similar to the untargeted attack case (cf. Fig. \ref{fig3}), where higher values of $k_-$ require a smaller fraction $p_c$ to destroy the coherence among the Kuramoto oscillators. This suggests that {\it increasing the repulsive coupling strength makes the synchronization more vulnerable, regardless of whether the attack is targeted or untargeted}.

% Indeed, it is an interesting observation that when a large fraction of nodes in the network are affected by repulsive coupling, a relatively smaller magnitude of the repulsive coupling strength is enough to disrupt the synchronization among the nodes. This phenomenon can be attributed to the collective influence of the nodes with repulsive coupling. When a significant portion of the network is affected, their combined effect becomes more pronounced and has a greater impact on the overall synchronization dynamics. In such cases, even a relatively weaker repulsive coupling strength can propagate through the network and disturb the coherence. The disruptive influence spreads more easily and effectively due to the network's interconnected nature and the mobile agents' random movement, leading to a breakdown of the collective synchronization behavior. This observation highlights the vulnerability of networks to targeted attacks when a substantial fraction of nodes experiences repulsive coupling. It emphasizes the importance of understanding the interplay between the fraction of affected nodes and the magnitude of repulsive coupling strength in determining the stability and synchronization properties of the network. By studying the behavior of synchronization in relation to the fraction of influenced nodes and the corresponding coupling strengths, we gain valuable insights into the robustness and vulnerability of networks under targeted attacks, enabling us to develop strategies for enhancing network security and robustness.

\subsubsection{Small world networks}

 Here, we explore the relationship between the critical negative coupling strength $k_-^c$ and the rewiring probability $q$ in small world networks. The degree distribution of small world networks is given by Eq.\ \eqref{26}, which is quite complex. Therefore, we could not obtain a closed-form solution for $k_-^c$ in this case. Instead, we adopt a numerical approach to calculate $k_-^c$. First, we numerically solve Eq.\ \eqref{30} to find the cutoff degree $d_c$. Then, we directly evaluate Eq.\ \eqref{32} to determine the critical repulsive coupling strength $k_-^c$.

\begin{figure}[ht]
	\centerline{\includegraphics[scale=0.5]{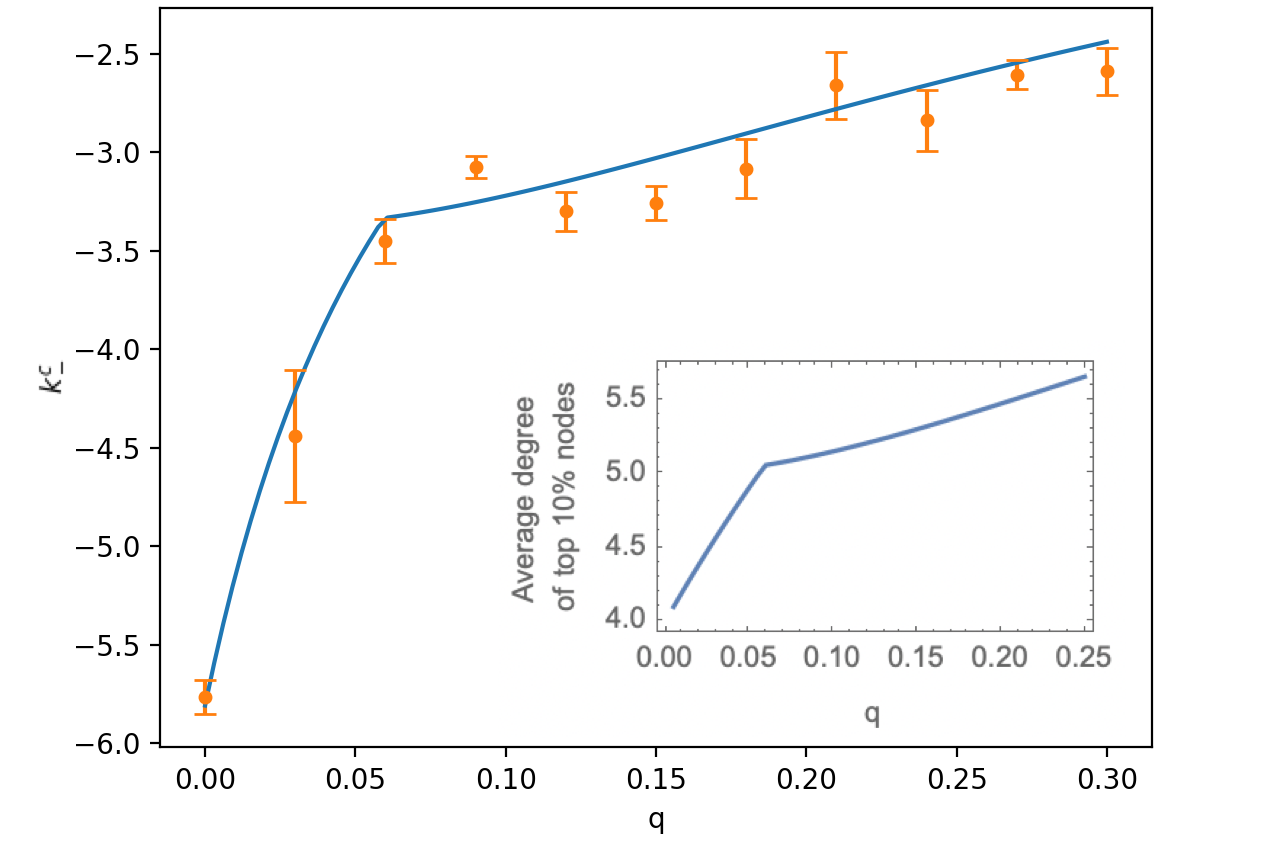}}
	\caption{{\bf Synchronizability of small world networks under targeted attacks}: As the rewiring probability $q$ increases, the strength of negative coupling $k_-^c$ needed to disrupt synchronization decreases. Thus, with more rewiring, a weaker negative coupling can disrupt the coherence among the nodes and lead to desynchronization. In the initial ring lattice structure, the network is regular, making it quite resilient under targeted attacks. However, as rewiring increases, the degree fluctuations grow, creating strategic targets and making the attack more effective. Other parameters: $n=100$, $K=2$, $p=0.1$, $k_+=1$, $N=500$, $\Delta \omega=1$, and $\Delta T=0.0001$. The inset shows the analytically computed average degree of top $10\%$ of most well connected nodes as a function of the rewiring probability $q$ for $K=2$ in the thermodynamic limit $n\rightarrow\infty$.
	}\label{fig7}
\end{figure}

 The results obtained from this numerical approach are plotted alongside the simulations in Fig. \ref{fig7}. We compare the values of $k_-$ obtained numerically with those obtained from Eqs.\ \eqref{30} and \eqref{32}, providing a visual representation of the agreement between theory and practice. %This figure allows us to validate our theoretical understanding of the system and confirms the accuracy of our approach in determining the critical repulsive coupling strength for small world networks. 
%The comparison in the figure demonstrates the agreement between the numerical solutions and simulations, further reinforcing the robustness of our theoretical framework and providing valuable insights into the behavior of small world networks under targeted attacks.
Figure  \ref{fig7} illustrates the impact of rewiring probability $q$ on the critical negative coupling strength $k_-^c$ required to disrupt synchronization in small-world networks. The parameters used in the simulations are $n=100$, $K=2$, $p=0.1$, $k_+=1$, $N=500$, $\Delta \omega=1$, and $\Delta T=0.0001$.
%As the rewiring probability $q$ increases, the network transitions from a regular lattice to a more random structure. In the regular ring-like lattice, each node has the same degree of $2K$ and is connected to its closest neighbors. However, as links are rewired with a higher probability, long-distance connections between previously unrelated (not directly connected) nodes are introduced, dramatically decreasing the network diameter. 
The figure shows that as the rewiring probability $q$ increases, the magnitude of the critical negative coupling strength $k_-^c$ decreases. This means that with a higher probability of rewiring, a relatively smaller magnitude of the negative coupling strength is sufficient to disrupt the coherence among the phase oscillators and lead to desynchronization.

Note, that increasing the rewiring probability $q$ showed the opposite effect on the system under targeted (Fig. \ref{fig4}) and untargeted (Fig. \ref{fig7}) attacks. If, during untargeted attacks, higher rewiring facilitated synchrony, in case of targeted attacks, it hindered synchronizability. This is because original lattice is completely uniform and thus regular, while rewiring introduces degree fluctuations that can be exploited during targeting.

We should also address the unexpected corner in Fig. \ref{fig7}, occurring at $q=0.06$. It is directly related to a very similar corner in the plot of the average degree of top $10\%$ highest degree nodes as a function of $q$ (see the analytic curve in the inset of Fig. \ref{fig7}). This is caused by the fact that for $q<0.06$ there are not enough nodes with degree $5$ and higher, so nodes with degree $4$ make the cutoff. For $q>0.06$ all the selected nodes have degree $5$ or higher. Thus below $q=0.06$, an infinitesimal increment of $q$ replaces degree $4$ nodes with higher degree nodes, whereas above  $q=0.06$, same increment of $q$ replaces degree $5$ nodes with higher degree nodes, resulting in discontinuously larger gain of average degree in the top $10\%$ of most well connected nodes.

% The underlying mechanism behind this observation lies in the interplay between local clustering and global connectivity in complex networks. In the initial regular lattice structure, nodes tend to form clusters due to high local clustering. Information and influence primarily propagate within these clusters, requiring a stronger negative coupling to disrupt synchronization. However, as the rewiring probability increases, the random connections introduced disrupt these clusters and decrease local clustering. This disrupts the flow of information within the network and enhances global connectivity. As a result, the magnitude of the critical negative coupling strength required to disrupt synchronization decreases, indicating that the network becomes more vulnerable to repulsive coupling. Overall, the figure demonstrates the sensitivity of small-world networks to changes in the rewiring probability. It highlights the delicate balance between local clustering and global connectivity and how it affects the synchronization dynamics. Understanding these dynamics is crucial for analyzing the robustness and vulnerability of small-world networks in the face of targeted attacks and disruptions.

\subsubsection{Scale-free networks}\label{Scale-free networks_targeted}

In the case of scale-free networks with degree distribution $\Pr(k)\propto k^{-\gamma}$, where $1<\gamma\leq 3$, the left-hand side of Eq.\ \eqref{13} diverges to negative infinity (since $\frac{\langle d^2 k\rangle}{\langle d\rangle^2}\rightarrow-\infty$), indicating absolute vulnerability to targeted attacks. In other words, any finite fraction of the most connected nodes being corrupted can disrupt synchronization, regardless of the magnitude of the negative coupling strength $k_-$. This contrasts with the untargeted attack scenario that are highly robust against untargeted attacks. Targeted attacks corrupt hubs whereas untargeted attacks corrupt   nodes at random, which are predominantly leafs and other low degree nodes. 

This observation aligns with earlier findings \cite{callaway2000network,watts2002simple,motter2002cascade,cohen2000resilience,cohen2001breakdown,albert2000error} on the structural robustness of heterogeneous networks. Heterogeneous networks, which include scale-free networks as a special case, are characterized by a wide range of node degrees. They are structurally robust against random node removal because the majority of nodes have low degrees and their removal does not significantly affect the overall connectivity. However, when we selectively remove important nodes, such as hubs, the network structure becomes fragmented, and its robustness is compromised \cite{newman2018networks,molloy1995critical}. 
%Similarly, in the context of synchronization dynamics, preferential removal or attack on important nodes (hubs) in scale-free networks leads to the fragmentation of the network and a drastic alteration in synchronization behavior. 
This fragility to preferential attacks on hubs is a consequence of the inherent structure of scale-free networks, where a small number of highly connected nodes play a crucial role in maintaining the overall connectivity and coherence. Therefore, {\it our findings highlight the dual nature of scale-free networks—they possess robustness against random disruptions but exhibit fragility when targeted attacks are directed towards hubs for $1<\gamma\leq3$}. These results resonate with the earlier studies \cite{callaway2000network,watts2002simple,motter2002cascade,cohen2000resilience,cohen2001breakdown,albert2000error} on the structural robustness and vulnerability of heterogeneous networks, emphasizing the intricate relationship between network topology, targeted attacks, and system dynamics.

\begin{figure}[ht]
	\centerline{\includegraphics[scale=0.5]{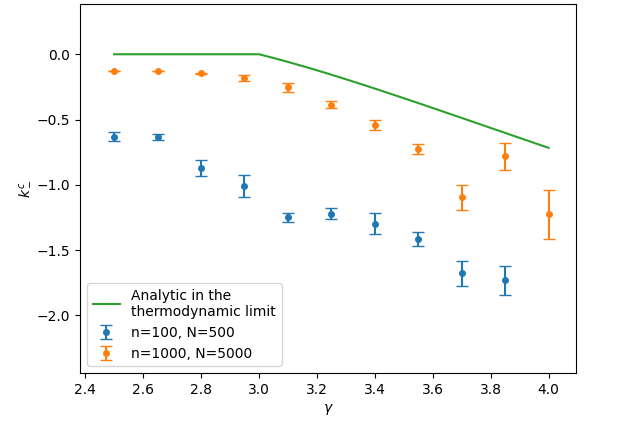}}
	\caption{{\bf Synchronizability of scale-free networks under targeted attacks}: The vulnerability of scale-free networks to targeted attacks depends on the network's size and the distribution of connections. Achieving a perfect match between theoretical predictions and simulations can be tricky due to sensitivity of scale-free networks to finite size effects, and computational constraints. Nonetheless, larger networks exhibit behavior closer to the analytical predictions, while smaller networks display more significant deviation. Hubs with limited-degree have a diminished ability to disrupt synchronization, demanding stronger negative couplings to achieve the same disruptive effect. Other parameters: $p=0.1$, $k_+=1$, $\Delta \omega=1$, and $\Delta T=0.0001$. 
	}\label{fig8}
\end{figure}

 For scale-free networks with $\gamma > 3$, we can explicitly compute the summation terms in Eq.\ \eqref{32} as shown in the following equation 

%\begin{widetext}
\begin{equation} \label{36}
	\begin{split}
		\sum_{d=d_c}^\infty  d^2\Pr(d)
		&=
		\sum_{d=d_c}^\infty  \frac{d^{2-\gamma}}{\zeta(\gamma)}
		=
		\frac{\zeta(\gamma-2, d_c)}{\zeta(\gamma)},
		 \\
		\sum_{d=1}^{d_c-1} d^2\Pr(d)
		&=
		\sum_{d=1}^{d_c-1} \frac{d^{2-\gamma}}{\zeta(\gamma)}
		\\
		&=\sum_{d=1}^\infty  \frac{d^{2-\gamma}}{\zeta(\gamma)} 
		-\sum_{d=d_c}^\infty  \frac{d^{2-\gamma}}{\zeta(\gamma)}
		\\
		&=\frac{\zeta(\gamma-2)-\zeta(\gamma-2,d_c)}{\zeta(\gamma)} .
	\end{split}
\end{equation}
%\end{widetext}
Here $\zeta(\cdot,\cdot)$ represents the Hurwitz zeta function. Combining this with Eqs.\ \eqref{28} and \eqref{32}, we get

%\begin{widetext}
\begin{equation} \label{37}
	\begin{split}
		k_-^c =
		-\Bigg(&
		k_+\bigg( \frac{\zeta(\gamma-2)}{\zeta(\gamma-2,d_c)}-1 \bigg)
		\\
		&-\frac{\zeta(\gamma-1)^2}{\zeta(\gamma)\zeta(\gamma-2,d_c)}
		\sqrt{\frac{8}{\pi}}\frac{n}{N}\Delta\omega
		\Bigg).
	\end{split}
\end{equation}
%\end{widetext}

Now we calculate the cutoff degree $d_c$ in terms of $p$. It must be chosen such that it separates the top $p$ fraction of nodes. In mathematical terms, this is expressed as
\begin{equation} \label{38}
	\begin{split}
		p 
		=
		\sum_{d=d_c}^{\infty} \Pr{(d)} 
		=
		\frac{1}{\zeta(\gamma)}\sum_{d=d_c}^{\infty} d^{-\gamma}
		=
		\frac{\zeta(\gamma, d_c)}{\zeta(\gamma)}.
	\end{split}
\end{equation}
Inverting this, we get
\begin{equation} \label{39}
	\begin{split}
		d_c 
		=
		\zeta^{-1}\left(\gamma,\zeta(\gamma)p\right),
	\end{split}
\end{equation}
where $\zeta^{-1}(x,y)$ denotes the inverse of the Hurwitz zeta function with $x$ fixed: $\zeta^{-1}(x,\zeta(x,y))=y$. 

 Figure \ref{fig8} presents the relationship between the critical negative coupling strength and the scale-free exponent $\gamma$. When it comes to targeted attacks on scale-free networks, the dynamics are highly influenced by the system's size, and our theoretical derivations are valid in the thermodynamic limit, i.e., for $n,N \to \infty$ and in the rapid movement limit of mobile agents, i.e., for $\Delta T \to 0$. Unfortunately, the simulations with scale-free networks are highly sensitive to finite size effects \cite{serafino2021true,boguna2004cut}, and due to the limited computation capacity, it is challenging to perfectly match numerical simulations with analytic predictions. However, as expected, increasing the network size $n$ and the number of mobile agents $N$ while maintaining a constant ratio $\frac{n}{N}$ brings the simulations closer to the thermodynamic limit and improves the agreement with analytic predictions.

 The finite size effects can be understood intuitively as follows. In scale-free networks, the degrees of the corrupted hubs are limited in finite-sized systems. As a consequence, these hubs have less influence on the synchronization dynamics compared to hubs in larger networks. In order to disrupt the synchronization, limited-degree hubs need to possess stronger negative couplings. This requirement arises because their reduced influence necessitates a more potent disruptive force to achieve incoherence.

 Overall, these findings highlight the intricate relationship between network size, topology, and the effectiveness of targeted attacks on scale-free networks. They remind us of the complex interplay between system properties, highlighting the importance of considering various factors when assessing the vulnerability of networks to targeted attacks.

\subsection{Targeted Attacks: Unveiling the Impact of Targeting Low-Degree Nodes}\label{Targeted Attacks: Unveiling the Impact of Targeting Low-Degree Nodes}

The mathematical results derived in the previous section can also be applied to a scenario where the low-degree nodes are targeted instead of the high-degree ones. In this case, we exchange the roles of $k_-$ and $k_+$ so that the nodes with the highest degrees now have positive couplings represented by $k_+$. We also substitute $p$ with $1-p$ while using $p$ to describe the fraction of disruptive nodes.

Under this type of targeted attack strategy, regular networks behave identically to the previous two cases. However, heterogeneous networks become even more robust compared to untargeted attacks. To understand this, we sort the nodes based on their degrees, where $d_\alpha$ represents a non-decreasing sequence. Since the highest degree nodes now have positive couplings $k_+$, the coupling strength $k_\alpha$ also exhibits a non-decreasing trend. By applying Chebyshev's sum inequality, we can establish the following relationship:

\begin{equation}\label{40}
	\frac{\langle d^2 k\rangle}{\langle d\rangle^2} \ge \frac{\langle d^2\rangle \langle k\rangle}{\langle d\rangle^2}.
\end{equation}

This inequality indicates that {\it the synchronization condition is more easily satisfied when lower-degree nodes are targeted compared to the untargeted case}. It further reinforces the enhanced robustness of the network under this targeted attack strategy.

\section{Exploring opinion dynamics: random walk of polarized agents}\label{Exploring opinion dynamics: random walk of polarized agents}
To showcase the generality and applicability of our results, we next consider a different type of internal dynamics and interactions: the cusp catastrophe model for polarization within and across the individuals \cite{van2020polarization} based on the Ising model of opinion. 
%then we incorporate it into our random walking agents model, derive the consensus condition, and validate it through simulations (Sec.\ \ref{Cusp catastrophe as internal dynamics of mobile agents}).

Let us first consider the internal dynamics of one individual as described in Ref.\ \cite{van2020polarization}. Each person forms their attitude about a subject matter based on an interconnected network of issues related to this subject. For example, if the subject is meat consumption, the issues could consist of beliefs (meat consumption doesn't affect climate), feelings (loves steak), and behavioral patterns (eats burgers). Each of these issues is treated as a binary node $x_i=-1,1$ indicating if the node label holds true for the given person. The overall opinion is given by an average over all subparts of attitude, i.e., network nodes (note, that the network is inside the individuals head, we are not discussing human-to-human interactions yet). The edge weights are given by $\omega_{ij}$. Additionally, one considers external influences affecting each issue $\tau_i$ (all their friends eat burgers) and attention to the subject matter $\mathscr{A}$ (how important the person thinks this topic is).

When the attention $\mathscr{A}$ is low, the connected issues can be misaligned $x_i\ne x_j$ (they can think that meat consumption affects the environment and still eat lots of burgers). However, as the person spends more and more time thinking about the topic, the cognitive dissonance tends to align the nodes with each-other and with the external influence. In other words, high attention implies the lower misalignment function

\begin{equation}\label{ising}
	\mathcal H=-\sum_{i}\tau_i x_i - \sum_{i,j}\omega_{ij}x_i x_j.
\end{equation}

This equation is known as the Ising model and is well studied in physics. The analogue of \textit{high attention} in opinion dynamics is \textit{low temperature} in the Ising model since both result in the lower value of Eq.\ \eqref{ising}. The overall opinion $\phi$ is analogous to the magnetization in the Ising model. Magnetization, on the other hand, has a cusp catastrophe behavior as a function of temperature and external influence in the Ising model. This can be directly translated to opinions: the opinion changes smoothly as a function of external influence $I$ for a low value of attention, while for high attention, hysteresis appears and, depending on the initial state, agent's opinion may be positive or negative for the same attention $\mathscr{A}$ and external influence $I$. The normal form dynamical equation describing a cusp catastrophe in its stationary states is given below:

\begin{equation}\label{cusp_dynamics_for_single}
	\dot\phi = f(\phi) = -\phi^3 + (\mathscr{A}-\mathscr{A}_c) \phi + I.
\end{equation}
Here $\phi$ stands for opinion, $\mathscr{A}$ indicates the attention to the subject matter, $\mathscr{A}_c$ stands for the critical value of attention beyond which the hysteresis appears, and $I$ describes the external influence coming from interactions with other individuals. For an in-depth study of this model, along with the description of interactions, and different real-world examples see Ref.\ \cite{van2020polarization}.

\subsection{Cusp catastrophe as internal dynamics of mobile agents}\label{Cusp catastrophe as internal dynamics of mobile agents}
The internal variable of mobile agents that stood for the phase $\phi_i\in\mathbb{S}^1$ will now be a real number $\phi_i\in\mathbb{R}$ denoting the opinion of the agent (note, that the result in Eq.\ \eqref{9} remains the same, in fact the internal state could even be a vector or a tensor). We consider the internal dynamics of agent $i$ to be given by Eq.\ \eqref{cusp_dynamics_for_single}

\begin{equation}\label{cusp_dynamics}
	\dot\phi_i = f(\phi_i) = -\phi_i^3 + (\mathscr{A}-\mathscr{A}_c) \phi_i + I.
\end{equation}
%Here $A$ stands for the individual's attention to the subject matter, and $I$ is the information bias or external influence the agent is exposed to. Equation \eqref{cusp_dynamics} predicts that for low values of attention $A<A_c$ the stationary opinion is a smooth function of $I$. While for high values of attention $A>A_c$, 

For the sake of simplicity, we consider that agents have a constant high value of attention $\mathscr{A}>\mathscr{A}_c$. And that the external influence experienced by each agent depends linearly on the neighbors' opinions. The coupling constants again vary between discussion venues through which the agents move randomly.
\begin{equation}
	\begin{split}
		&\dot{\phi}_i = F(\phi_i)+k_{\alpha} \sum_{j \in O_{\alpha}} H(\phi_i,\phi_j),
		\\
		&F(\phi_i) =  -\phi_i^3 + (\mathscr{A}-\mathscr{A}_c) \phi_i,
		\\
		&H(\phi_i, \phi_j) = \phi_j.
	\end{split}
\end{equation}
Here the interaction term $H(\phi_i, \phi_j)$ ensures that agents with positive opinions affect their neighbors in the positive direction proportional to their conviction level (as long as the coupling $k_\alpha$ is positive). For friendly, constructive discussions $k_\alpha$ will be positive, meaning that the listener takes the speakers words at the face value. For antagonistic interactions the coupling may well be negative, indicating that the listener will want to distance themselves from the speaker. 

Employing Eq.\ \eqref{9}, we can write down the state equations in the weak coupling limit

\begin{equation}\label{MFT_cusp}
	\begin{split}
		&\dot{\phi}_i = -\phi_i^3 + (\mathscr{A}-\mathscr{A}_c) \phi_i+ \frac{\tilde k}{N}  \sum_{j=1}^{N} \phi_j,
		\\
		&\tilde k = \dfrac{N}{n} \dfrac{\langle d^2 k \rangle}{ {\langle d \rangle}^2}.
	\end{split}
\end{equation}

\begin{figure}[ht]
	\centerline{\includegraphics[scale=0.5]{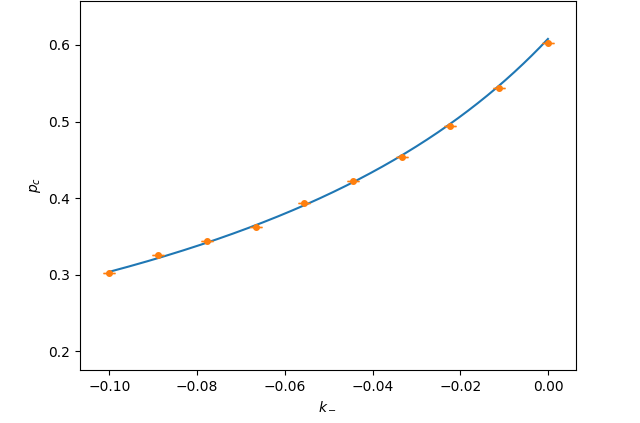}}
	\caption{{\bf Emergence of consensus in regular networks under untargeted attacks}: Critical fraction of disruptive nodes $p_c$ as a function of their coupling strength $k_-$. $p_c$ is the fraction of disruptors in the network, beyond which consensus becomes impossible. Other parameters: $q=0.1$, $(\mathscr{A}-\mathscr{A}_c)=1$, $k_+=0.1$, and $\Delta T=0.001$.
	}\label{fig9}
\end{figure}

As expected, the result is of the same form as Eq.\ \eqref{cusp_dynamics}, but for globally coupled agents. We will initiate the agents with polarized opinions. If the effective coupling $\tilde k$ is large, the agents will achieve a consensus, whereas for low values of $\tilde k$ the agents will remain polarized. We can find the critical effective coupling necessary for the consensus using bifurcation analysis. Treating $(\mathscr{A}-\mathscr{A}_c)$ as a positive constant, and $I$ as a parameter in Eq.\ \eqref{cusp_dynamics}, we evaluate the bifurcation conditions $f(\phi_i)=0$ and $f'(\phi_i)=0$ to get the bifurcation curve

\begin{equation} \label{bifurcation_curve}
	I = \pm \frac{2(\mathscr{A}-\mathscr{A}_c)^{\frac{3}{2}}}{3\sqrt{3}}.
\end{equation}
Without loosing generality, we focus on the positive solution and compute the two equilibrium points for opinion
\begin{equation}
	\begin{split}
		\phi^- &= -\sqrt{\frac{\mathscr{A}-\mathscr{A}_c}{3}},
		\\
		\phi^+ &= 2\sqrt{\frac{\mathscr{A}-\mathscr{A}_c}{3}}.
	\end{split}
\end{equation}
This, in turn, helps us calculate the interaction term $I$. Let us assume that in the initial state the opinions are divided into fraction $q$ that has a negative opinion and the rest $(1-q)$ tat that thinks positively. Then the influence of population opinions on each individual is

\begin{equation}\label{MFT_interaction}
	\begin{split}
		I &= \frac{\tilde k}{N}  \sum_{j=1}^{N} \phi_j = \tilde k \big(q\phi^-+(1-q)\phi^+\big)
		\\ & = 
		\tilde k\sqrt{\frac{\mathscr{A}-\mathscr{A}_c}{3}}(2-3q).
	\end{split}
\end{equation}

The consensus appears when the interaction term Eq.\ \eqref{MFT_interaction} exceeds the bifurcation value Eq.\ \eqref{bifurcation_curve}. This yields the condition

\begin{equation} \label{globally_coupled_consensus_condition}
	\tilde k > \frac{2(\mathscr{A}-\mathscr{A}_c)}{6-9q}.
\end{equation}
 
Note that since we considered only the positive solution for the bifurcation curve, Eq.\ \eqref{globally_coupled_consensus_condition} is relevant only when positive opinion prevails, i.e., $q>0.5$. For the reversed scenario, the symmetry of the problem implies that one simply needs to replace $q$ by $1-q$ in Eq.\ \eqref{globally_coupled_consensus_condition}. This condition for consensus is for the globally coupled system Eq.\ \eqref{MFT_cusp}. Now we can use the expression for $\tilde k$ to arrive at the general consensus condition for the random walking agents

\begin{equation} \label{cusp_general_condition}
	 \dfrac{\langle d^2 k \rangle}{ {\langle d \rangle}^2} > \dfrac{n}{N}\frac{2(\mathscr{A}-\mathscr{A}_c)}{6-9q}.
\end{equation}

Equation \eqref{9} predicts that the impact of the network topology and the coupling distribution (or the attack strategy) remains independent of the internal dynamics of the agents (compare Eq.\ \eqref{cusp_general_condition} with Eq.\ \eqref{13}). In order to avoid redundancy, we only present the numerical experiments with a regular network under untargeted attacks.

The consensus condition for untargeted corruption of a fraction $p$ of discussion venues is given by a derivation identical to Eq.\ \eqref{21}

\begin{equation}\label{pc_for_cusp}
	p_c=\dfrac{1}{k_{+}-k_{-}} \left(k_{+}-\frac{2(\mathscr{A}-\mathscr{A}_c)}{6-9q}\frac{n}{N} \frac{{\langle d \rangle}^2}{\langle d^2 \rangle} \right).
\end{equation}

Figure \ref{fig9} shows the numerical validation of Eq.\ \eqref{pc_for_cusp} with a random 3-regular network of $n=100$ nodes and $N=1000$ agents. The initial split of opinions $q=0.1$, the attention $(\mathscr{A}-\mathscr{A}_c)=1$, positive coupling $k_+=0.1$, and the time intervals $\Delta T=0.001$.

\section{DISCUSSION AND CONCLUSIONS} \label{DISCUSSION}

In this study, we have explored the dynamics of mobile agents as they navigate a complex network and interact with diverse sets of neighbors in different environments during their movements. The agents' internal dynamics are described by arbitrary first-order differential equations, and their interactions are governed by an arbitrary function of agent states. The network nodes act as interaction venues for the agents and exhibit heterogeneity. This variation between the nodes is modeled by a parameter known as the coupling constant, which regulates the strength of interactions between agents within the node.

Our analytic framework is validated in two distinct scenarios motivated by different applications. In both cases, we consider individuals moving through a network of interaction venues, including offices, bars, online chats, social media comments sections, news articles, and other physical or digital locations. The internal dynamics and interactions of individuals vary between the two applications. 

The first application is synchronization of brain activity among groups of interacting individuals. This phenomenon has been observed in various recent experiments, employing techniques such as MRI, EEG, and eye tracking \cite{wheatley2019beyond, hu2017brain, perez2017brain,czeszumski2020hyperscanning, sievers2020consensus, wohltjen2021eye}. We model this behavior by representing agents as Kuramoto oscillators, which tend to synchronize upon interactions, and explore the global synchronizability of the system.

The second application pertains to a cusp catastrophe model of opinion dynamics \cite{van2020polarization}. This recently published model delves into polarization within and across individuals, highlighting how Ising-like interactions between related issues lead to the emergence of hysteresis in opinion dynamics when attention to the subject matter is high. We incorporate this cusp catastrophe model as the internal dynamics in our random-walking model to examine the possibility of consensus.

Based on our analysis, which becomes exact in the limit of weak couplings, we derive effective differential equations governing the evolution of agents' internal states Eq.\ \eqref{9}. Our analysis accounts for the network topology and coupling heterogeneity, incorporated in the expression for the effective coupling. Through our analytical findings we can make several general observations: small networks with many agents facilitate a strong effective coupling, high-degree nodes exert a strong influence on the system behavior, and node degree fluctuations play a crucial role in stimulating interactions. %The latter phenomenon aligns with the concept of converse symmetry breaking \cite{nishikawa2016symmetric}: uniformity in node degrees leads to weak coupling of internal states, while degree fluctuations promote cohesion among agents. 
Additionally, we find that designing the network structure intelligently, with inherent node variations in mind, can improve its functionality. In particular, aligning node degrees with their respective couplings enhances interactions.

An important strength of our analysis lies in its ability to accommodate diverse network nodes, both in terms of their degrees and their internal dynamics. The coupling constants associated with nodes can be arbitrarily distributed, enabling us to explore the interplay of positive and negative couplings. Nodes with negative couplings can be interpreted as disruptors of the system. Moreover, our approach allows for the selection of disruptive nodes to be dependent on the network topology, leading to different attack strategies and notions of robustness concerning these attacks.

To investigate the effect of different network topologies on facilitating coherence under disruptive influences, we consider two distinct methods of introducing nodes with negative couplings. First, we randomly select a portion of nodes and assign them negative coupling strengths. Second, we employ a more sophisticated approach by specifically targeting high-degree nodes and analyzing the consequences of this preferential placement.

Through detailed analysis, we provide analytical proof that under untargeted attacks, scale-free networks with a power-law exponent ($\gamma$) between $1$ and $3$ exhibit the highest robustness, while regular networks are the most vulnerable. Random networks fall somewhere in between these extremes. However, when attackers strategically target high-degree nodes, the response of the system changes. Scale-free networks with $1<\gamma\leq3$ become the weakest, their coherence easily disrupted even with mildly negative coupling strengths ($k_-$). This reversal is well illustrated in small world networks too, where increasing the rewiring probability makes coherence easier under untargeted attacks but harder under targeted attacks. The networks that were previously the most robust under untargeted attacks now become the most susceptible when targeted. This finding aligns with previous studies exploring complex networks' structural robustness \cite{albert2000error}. Regular networks, which were initially vulnerable, show higher robustness under targeted attacks than heterogeneous networks. However, %the specific details of the dynamics, such as agent frequency distribution, coupling strengths, and the fraction of disruptors, influence the robustness of regular networks. 
it is important to note that the most robust network topology under targeted attacks is not universally fixed and depends on various factors. The heterogeneity of node degrees in the network plays a crucial role. On one hand, heterogeneity promotes coherence when it comes to nodes with positive coupling. However, this very heterogeneity also creates potential targets for disruptors aiming to destabilize the network.

Our analysis indicates that achieving coherence becomes more challenging when targeted attacks are directed toward higher-degree nodes, demonstrating a decreased network robustness. Additionally, we investigate the effects of targeting lower-degree nodes with repulsive couplings and find that the coherence conditions are more readily satisfied under this preferential attack. In addition to formulating a comprehensive analytical solution, we supplement our research with extensive numerical experiments to corroborate our discoveries. While the majority of these outcomes reveal robust concurrence with our analytical conclusions, it becomes evident that scale-free networks exhibit pronounced finite-size effects. In conclusion, the relationship between network topology and internal coherence is complicated. The most robust network structure under targeted attacks depends on agents' internal dynamics, interaction function, coupling strengths, and the proportion of disruptors. Gaining a deep understanding of these intricate details will enable us to identify the network structures that are most robust when faced with strategic attacks. By unraveling this interplay, we can enhance our ability to design robust networks capable of withstanding and recovering from disruptions.

\section*{Acknowledgments} G.M. and R.M.D. express sincere gratitude for the support provided by Army Research Award number W911NF-23-1-0087. S.N.C. and A.H. would like to acknowledge the support of the National Science Foundation under Grant No.\ 1840221.

\typeout{}
\bibliographystyle{apsrev4-1}
\bibliography{mobile_agents_on_networks}

\end{document}